\documentclass[sn-nature,iicol]{sn-jnl}% Style for submissions to Nature Portfolio journals
\usepackage{pifont}

\jyear{2024}%

\theoremstyle{thmstyleone}%
\theoremstyle{thmstyletwo}%

\theoremstyle{thmstylethree}%

\begin{document}
\title[Article Title]{Mechanical properties of crystalline--amorphous composites: generalisation of Hall--Petch and inverse Hall--Petch behaviours }

\author{\sur{Zhibin} \fnm{Xu}}\email{zxucb@connect.ust.hk}
\author{\sur{Mengmeng} \fnm{Li}}\email{mlicc@connect.ust.hk}
\author*{\sur{Yilong} \fnm{Han}}\email{yilong@ust.hk}
\affil{\orgdiv{Department of Physics}, \orgname{The Hong Kong University of Science and Technology}, \orgaddress{\street{Clear Water Bay}, \city{Hong Kong}}}

\makeatletter
\patchcmd{\@maketitle}{\artauthors}{\centerline\artauthors}{}{}
\makeatother
%<150 words
\abstract{The strength, $\sigma_{\rm y}$, of a polycrystal decreases with mean grain diameter $D$ at $D\gtrsim50$ atoms (i.e. Hall--Petch behaviour) and increases at $D\lesssim50$ (i.e. inverse Hall--Petch behaviour). Our simulations generalise $\sigma_{\rm y}(D)$ to $\sigma_{\rm y}(D,l)$, where $l$ is the mean thickness of grain boundaries. For various particle compositions, the maximum strength is reached at $(D,l)\simeq(50, 6)$ particles for single-component face-centred-cubic solids and at $(D,l)\simeq(50, 2)$ for bidispersed or body-centred-cubic solids because of the different activation stresses of dislocation motions. The results explain recent alloy experiments and provide a way to exceed the maximum strength of polycrystals. Ductility and elastic moduli are also measured in the broad $(D,l)$ space. The regimes without a strength--ductility trade-off, the maximum ductility and ductile--brittle transitions are identified. These results obtained in $(D,l)$ space are important in solid mechanics and can guide the fabrication of crystalline--amorphous composites with outstanding mechanical properties.}


\maketitle

\section*{Introduction}\label{sec1}
Polycrystalline materials, such as ceramics, metals and alloys, are ubiquitous in nature and industries. Their strength or yield stress, $\sigma_{\rm y}$, increases as mean grain diameter $D$ decreases, i.e., the famous Hall--Petch (HP) behaviour discovered in the 1950s~\cite{hall1951deformation,petch1953cleavage}. This trend reverses at $D\lesssim10-20~{\rm nm}\simeq 50$ atoms, i.e., inverse Hall-Petch (IHP) behaviour~\cite{chokshi1989validity,trelewicz2007hall}. HP and IHP behaviours generally hold in all atomic and molecular polycrystals and have been intensively studied~\cite{pande2009nanomechanics,armstrong201460,cordero2016six,naik2020hall}. Aside from tuning $D$, a polycrystal's strength can also be enhanced by increasing pressure~\cite{zhou2020high} and decreasing temperature~\cite{naik2020hall} or the initial dislocation density~\cite{naik2020hall}. However, previous studies have focused on solid strengthening instead of generalising HP and IHP behaviours by tuning structural parameters in a broad range. An important simulation study obtained a non-monotonic $\sigma_{\rm y}$ by adding twin boundaries~\cite{li2010dislocation}; such $\sigma_{\rm y}$ is essentially covered by the HP and IHP behaviours because twin boundaries can be viewed as a special type of grain boundaries (GBs). Here we generalise the HP and IHP behaviours of $\sigma_{\rm y}(D)$ for polycrystals with mean GB thickness $l\simeq2$ particles to $\sigma_{\rm y}(D,l)$ by tuning structural parameters $D$ and $l$. 

\begin{figure*}[t]%
\centering
\includegraphics[width=\textwidth]{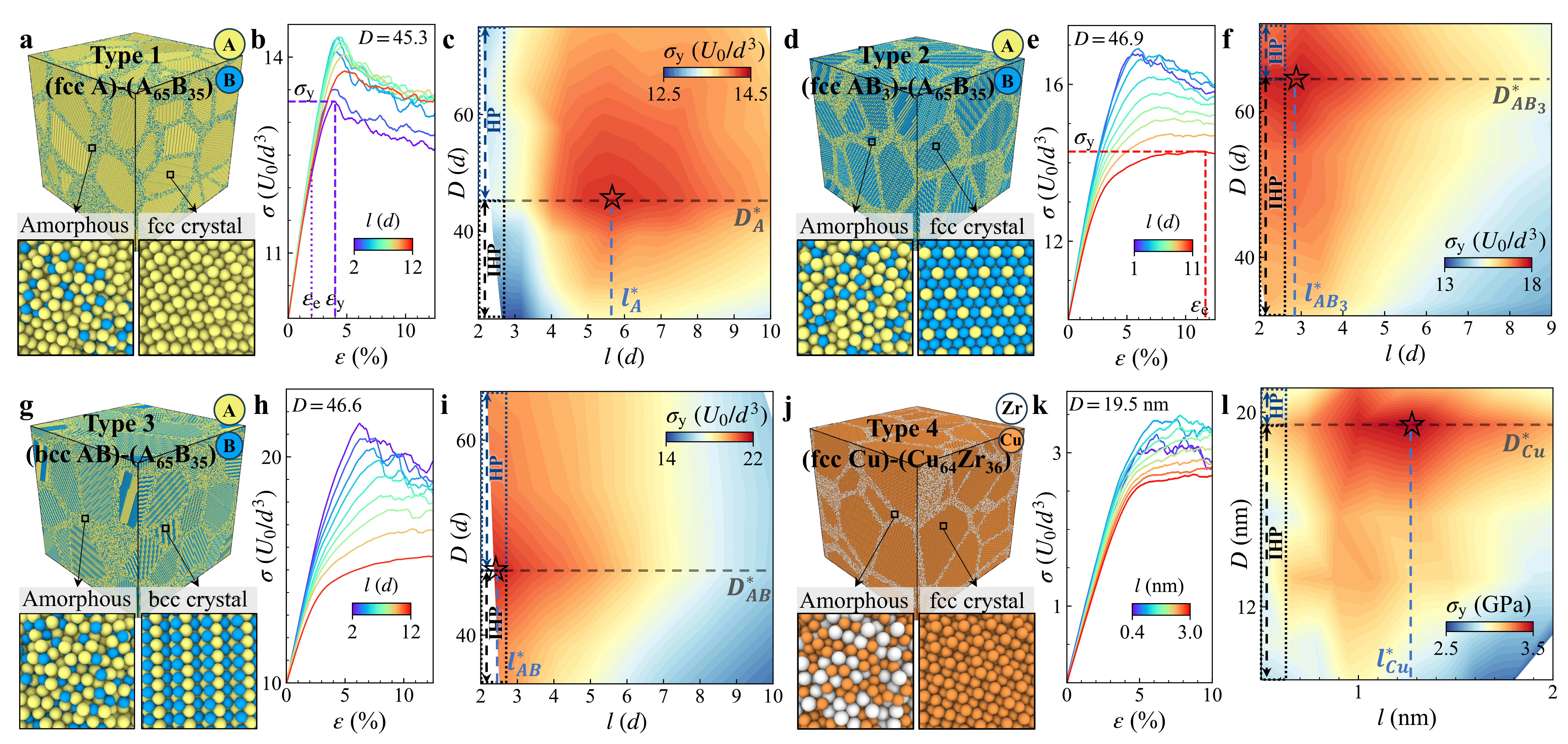}
\caption{Four types of crystalline--amorphous composites (a, d, g, j), their corresponding engineering stress--strain curves (b, e, h, k), and generalised HP and IHP behaviours in $(D,l)$ space (c, f, i, l). (a) (fcc-A)-(A$_{65}$B$_{35}$) denotes 100\% large particles in fcc crystalline regions and a 65:35 mixture of large and small particles in amorphous regions. (b) Stress--strain curve $\sigma(\varepsilon)$ of samples with a fixed $D=45.3$ and various $l$. Elastic limit $\varepsilon_{\rm e}$ represents the end of the linear regime. The maximum stress $\sigma_{\rm y}$ of each curve is used in (c). (c) Contour map of $\sigma_{\rm y}(D,l)$. The blue and black boxes at $l\simeq2$ are the conventional HP and IHP regimes, respectively. The horizontal dashed line represents the HP/IHP boundary at different $l$. The vertical dashed line represents GB thickness $l^*$, with the maximum yield stress $\sigma_{\rm y}^{\rm max}$ labelled as $\medwhitestar$. Similar to (a--c), the results of the three other types of systems are shown in (d-f) for (fcc-AB$_{3}$)-(A$_{65}$B$_{35}$), (g--i) for (bcc-AB)-(A$_{65}$B$_{35}$) and (j--l) for (fcc-Cu)-(Cu$_{64}$Zr$_{36}$) composites.}\label{fig1}
\end{figure*}

Thick-GB polycrystals are difficult to fabricate. Thus, the effect of $l$ on material properties remains unclear and poorly explored. In recent years, thick-GB polycrystals have been fabricated in glass-ceramics~\cite{holand2019glass} and alloys which are often called crystalline--amorphous composites~\cite{li2019amorphous}, dual-phase crystal--glass materials~\cite{gu2024phase}, or GB complexions~\cite{cantwell2014grain}. Such materials exhibit ultrahigh strength~\cite{holand2019glass,wu2017dual,ding2019thick}, superplasticity~\cite{su2021high},  ultralong fatigue life~\cite{hua2021nanocomposite}, hydrogen storage ability~\cite{li2019amorphous} and wear resistance~\cite{holand2019glass,li2019amorphous}. They are often fabricated by complex trial-and-error processes, so the size of crystalline and amorphous regions is difficult to control~\cite{gu2024phase} when studying the effect of $l$ on mechanical properties. Simulations of the effects of thick-GB on material properties are limited to Cu-based alloys within a narrow range of $l$~\cite{xiao2021mitigating,brink2018metallic}. A recent simulation measured the strength of 2D polycrystals with thick GBs; such solids only exhibited IHP behaviour, and the maximum strength at HP--IHP transition was not determined ~\cite{xu2023generalization}. Here, we systematically measure the strength of 3D solids so that both the HP and IHP behaviours can be generalised to the $(D,l)$ parameter space. We find that face-centred-cubic (fcc) and body-centred-cubic (bcc) lattices exhibit different behaviours because of their different activation stresses of dislocation motions. Aside from $\sigma_{\rm y}$, the fracture behaviours and elastic moduli in broad $(D,l)$ space are also studied, which has not been done before.

\section*{Generalisation of HP and IHP behaviours}\label{sec2}
We systematically vary $D$ and $l$ in four types of systems (Fig.~\ref{fig1}a,d,g,j) and measure their $\sigma_{\rm y}(D,l)$ by molecular dynamics simulations. Types 1--3 systems are composed of binary-sized particles, with the Lennard--Jones (LJ) pair potential $U(r)=4U_0[({d}/{r})^{12}-({d}/{r})^6]$ shifted to zero at $r>r_\textrm{c}=2.5d$. $d_{\rm AA}=1$ for large (A-type) particles; $d_{\rm BB}=0.88$ for small (B-type) particles; and $d_{\rm AB}/d_{\rm AA}=0.8$ following the Kob--Anderson binary LJ mixture model~\cite{kob1995testing}. In amorphous GBs, the number ratio between A and B particles is set as 65:35 because large A and small B particles can mix well without phase separation~\cite{bruning2008glass}. In crystalline grains, they are set as 100:0, 25:75 and 50:50 for Types 1--3 systems, respectively. Type 4 systems comprise pure Cu in crystalline grains and a 64:36 mixture of Cu and Zr atoms in GBs. All the number ratios above commonly exist in real alloys~\cite{porter2009phase}. The Cu-Zr mixture is a prototype of metallic glass~\cite{schuh2007mechanical}. The atomic interactions are described by embedded-atom-method potentials~\cite{mendelev2019development}. Crystalline regions are fcc for Types 1, 2 and 4 systems (Fig.~\ref{fig1}a,d,j) and bcc for Type  3 systems (Fig.~\ref{fig1}g). Uniaxial compression with a total strain of $\varepsilon=12.5\%$ and $\varepsilon=10\%$ is applied on LJ systems and Cu-Zr systems, respectively, all of which exceed their yielding points. Additional simulation details are given in Supplementary Information Section 1. 

\begin{figure*}[h]%
\centering
\includegraphics[width=\textwidth]{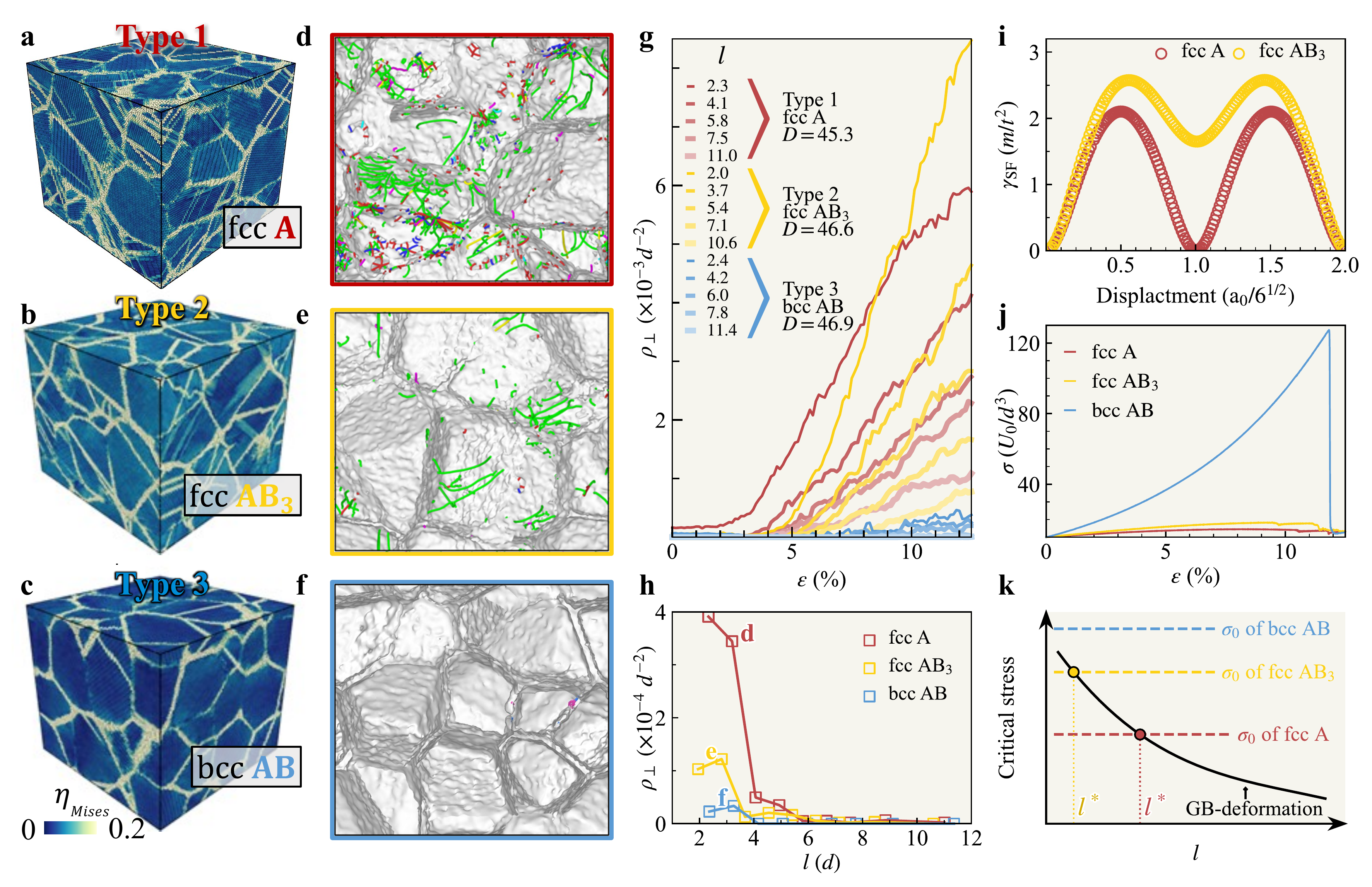}
\caption{Lattice symmetry and composition effects on the plastic deformation mechanism of crystalline--amorphous composites. (a--c) Von Mises shear strain, $\eta_{_{\rm{Mises}}}$ of Types 1--3 systems (Fig.~\ref{fig1}a,d,g) under $\varepsilon=3.18\%, 3.26\%$ and 4.50\% at $(D,l)=(45.3,5.8)$, $(46.9,2.8)$ and $(46.6,2.4)$, respectively, corresponding to their $\sigma_{\rm y}^{\rm max}$ labelled with $\medwhitestar$ in Fig.~\ref{fig1}c,f,i. (d--f) Dislocation lines in the whole 3D samples of (a--c).  Green: Shockley; pink: stair-rod; blue: full dislocation; red: other types of dislocations; grey: surfaces of grains. They are all projected to a 2D plane.  Defect formation processes in 3D are shown in Supplementary Videos 1--3. (g) Dislocation density $\rho_\perp$ in crystalline regions of Types 1--3 systems during loading. The details of $\rho_\perp$ are in Supplementary Information Section 3.2. (h) $\rho_\perp$ at the midpoint between $\varepsilon_{\rm e}$ and $\varepsilon_{\rm y}$ in Fig.~\ref{fig1}b,e,h. (i) Stacking fault energy $\gamma_{\rm sf}$ as a function of the displacement between two adjacent $\{111\}$ planes (Supplementary Information Fig.~S10b)~\cite{van2004stacking} in Types 1 and 2 fcc single crystals. $a_{0}$ is the lattice constant. (j) Stress-strain curves of three defect-free single crystals with compositions A, AB and AB$_3$ under uniaxial compression along the $\left[111\right]$ direction. The bcc crystal is much harder than fcc crystals and exhibits a sharp decrease at $\varepsilon=12\%$ (i.e. fracture).  For these defect-free crystals, the fracture corresponds to the formation of the first dislocation~\cite{salehinia2014crystal}. (k) Schematic of the required stress $\sigma_0$ for plastic deformation in crystalline grains and GBs.}\label{fig2}
\end{figure*}

The stress--strain curves $\sigma(\varepsilon)$ around the HP--IHP crossover are shown in Fig.~\ref{fig1}b,e,h,k. Each curve is averaged over three samples compressed along $x$, $y$ and $z$ directions for sufficient statistics. Both the maximum stress (i.e. yield stress $\sigma_{\rm y}$) and flow stress $\sigma_{\rm f}$ given by the plateau height of $\sigma(\varepsilon)$ have been commonly used as the material strength, and they exhibit similar HP/IHP behaviours~\cite{naik2020hall}. However, some samples exhibit strain softening without a plateau in $\sigma(\varepsilon)$. Thus, we show HP/IHP behaviours in terms of $\sigma_{\rm y}$ in the main text and in terms of the ambiguous $\sigma_{\rm f}$ in Supplementary Information Fig.~S5. 

The contour maps of $\sigma_{\rm y}(D,l)$ in Fig.~\ref{fig1}c,f,i,l can be viewed as the generalisation of the HP and IHP behaviours of $\sigma_{\rm y}(D)$. Conventional HP and IHP behaviours of $\sigma_{\rm y}(D)$ are well reproduced in thin-GB polycrystals ($l\simeq 2$). The presence of triple junctions increases $l$ to about 2 for polycrystals. The maximum of $\sigma_{\rm y}(D,l\!=\!2)$ is observed at $D^*\simeq50$ particles in Fig.~\ref{fig1}c,f,i,l which agrees well with the conventional HP--IHP crossover regimes of atomic polycrystals~\cite{naik2020hall}. $D^*$ remains at about 50 particles on the generalised HP--IHP boundary under different $l$ values (horizontal dashed lines in Fig.~\ref{fig1}c,f,i,l). Similar to $\sigma_{\rm y}(D)$, $\sigma_{\rm y}(l)$ is also non-monotonic for Types 1 and 4 systems. The maximum $\sigma_{\rm y}$ is at $(D^*,l^*)=(45.3,5.8)$ for Type 1 systems ($\medwhitestar$ in Fig.~\ref{fig1}c) and at $(D^*,l^*)=(19.5,1.5)~\textrm{nm}=(60,5)$ Cu atoms for type 4 systems ($\medwhitestar$ in Fig.~\ref{fig1}l), indicating that $(D^*,l^*)$ slightly depends on particle interaction. 

The deformations of polycrystals in conventional HP and IHP regimes occur via dislocation motions in crystalline grains and via deformations in GBs, respectively~\cite{pande2009nanomechanics}. Small grains produce few dislocation pile-ups on GBs, thus requiring high applied stress to initiate plastic flow in crystalline grains; meanwhile, the presence of many GBs results in low required stress for GB deformation~\cite{armstrong201460,naik2020hall}. The competition between the required stresses for the two deformation mechanisms leads to non-monotonic $\sigma_{\rm y}(D)$~\cite{pande2009nanomechanics}. 

The above conventional mechanisms of the HP and IHP behaviours of $\sigma_{\rm y}(D)$ for polycrystals with $l\simeq 2$ can similarly explain $\sigma_{\rm y}(l)$ in Fig.~\ref{fig1}c,f,i,l. In the $l<l^*$ regime, thick GBs reduce the generation and motion of dislocations~\cite{ding2019thick}, leading to few dislocation pile-ups on GBs (Fig.~\ref{fig2}, Supplementary Videos 1,2). Meanwhile, thick GBs act as a strong sink for absorbing dislocations~\cite{wang2007ductile}. Both effects cause a low stress concentration on thick GBs, so high applied stress is required for plastic deformation. Such dislocation-motion-controlled deformation decreases (i.e. strength increases) as $l$ increases (Supplementary Information Fig.~S11). In the $l>l^*$ regime, GB deformations dominate plastic deformation (Fig.~\ref{fig2}, Supplementary Videos 1,2), which agrees with previous simulations of crystalline--amorphous composites~\cite{xiao2021mitigating,xu2023generalization}. This increase in GB deformation (i.e. decrease in strength) as $l$ increases is consistent with the behaviours of glass nanopillars~\cite{jang2010transition,schuh2007mechanical2}. The competition between dislocation-motion-controlled and GB-deformation-controlled strengths causes a peak in $\sigma_{\rm y}(l)$, as sketched in Supplementary Information Fig.~S11. 

In contrast to the non-monotonic $\sigma_{\rm y}(l)$ at a fixed $D$ in Fig.~\ref{fig1}c,l for Types 1 and 4 systems, $\sigma_{\rm y}(l)$ in Fig.~\ref{fig1}f,i decreases almost monotonically with the maxima at $(D^*,l^*)=(64.4, 2.8)$ for Type 2 systems and at $(46.6, 2.4)$ for Type 3 systems. This difference amongst systems indicates that $\sigma_{\rm y}(l)$ depends on crystal composition and lattice symmetry.  We attribute the narrow regime of $l<l^*$ to the higher friction stress of moving dislocations in binary-composition fcc (Type 2) and bcc (Type 3) systems than in monodispersed-fcc (Types 1 and 4) systems.  This mechanism is shown in  Fig.~\ref{fig2}a-j and summarized in Fig.~\ref{fig2}k. As sketched in Fig.~\ref{fig2}k, the minimum stress to deform GB monotonically decreases with $l$~\cite{zhou2020high}, whereas the minimum stress $\sigma_0$ required to move a dislocation should be independent of $l$~\cite{zhu2012deformation}. When the applied stress exceeds one of the two minimum required stresses above, one type of plastic deformation occurs and largely pre-empts the other type of deformation. Thus, the intersection of the two stresses in Fig.~\ref{fig2}k gives $l^*$.  Given that dislocations require high stress to move in bcc or binary crystalline grains ($\sigma_{\rm 0}^{\rm bcc-AB}\gg\sigma_{\rm 0}^ {\rm fcc-AB_{3}}>\sigma_{\rm 0}^{\rm fcc-A}$), their $l^*$ in Fig.~\ref{fig1}f,i is smaller than that of monodispersed fcc grains (Fig.~\ref{fig1}e,l). 

First, we compare Types 1 and 2 fcc systems. Local deformation can be described by von Mises shear strain $\eta_{_{\rm Mises}}$ defined in Supplementary Information Eq.~S1. Particles with $\eta_{_{\rm Mises}}>0.12$ are set as plastic deformation regions~\cite{zepeda2017probing}, which exist in grains by dislocation motions and in GBs by shear-transformation zones (Fig.~\ref{fig2}a,b, Supplementary Video 1) at a small $l$. The dislocation lines in Fig.~\ref{fig2}d,e and dislocation density $\rho_\perp$ in Fig.~\ref{fig2}g show that numerous dislocations are generated in the Type 1 system during yielding. $\rho_\perp$ decreases with $l$ in Fig.~\ref{fig2}g, indicating that thick GBs suppress the formation of dislocations. For Types 1--3 systems, $\rho_\perp$ before reaching $\sigma_{\rm y}$ is nearly zero (Fig.~\ref{fig2}h) for samples with $l\geqslant l^*$ in Fig.~\ref{fig1}c,f,i, indicating that the plastic deformation is solely from GBs. Moreover, we observe that the dislocations in fcc systems are mainly 1/6$\langle112\rangle$ (Shockley) partial dislocations (Fig.~\ref{fig2}d,e) which require lower stress to move compared with other types of dislocations~\cite{hull2011introduction}. They are often emitted from crystal--amorphous interfaces and propagate into crystalline regions (Supplementary Video 1). For fcc crystals, the minimum shear stress for creating partial dislocation, $\sigma_{\rm 0}$, linearly increases with stacking fault energy $\gamma_{\rm sf}$~\cite{zhu2012deformation}. We measure $\gamma_{\rm sf}$ by the generated stacking fault method \cite{van2004stacking} (Supplementary Information Fig.~S10) and obtain $\gamma_{\rm sf}^{\rm fcc-AB_{3}}>\gamma_{\rm sf}^{\rm fcc-A}$ (Fig.~\ref{fig2}i). Thus, $\sigma_{\rm 0}^{\rm fcc-AB_{3}}>\sigma_{\rm 0}^{\rm fcc-A}$, and Type 2 (fcc-AB$_3$) systems have fewer dislocations than Type 1 (fcc-A) systems, which is in accordance with Fig.~\ref{fig2}d,e and Supplementary Videos 1, 2. Moreover, unlike in Type 1 systems whose dislocations appear immediately after the onset of plastic deformation (i.e. $\varepsilon>\varepsilon_{\rm e}$ in Fig.~\ref{fig1}b), $\rho_\perp$ of Type 2 systems is nearly zero (Fig.~\ref{fig2}g) during the strain hardening stage ($\varepsilon_{\rm e}<\varepsilon<\varepsilon_{\rm y}$ in Fig.~\ref{fig1}e, Supplementary Video 2). Thus, dislocations barely contribute to the yield at $\varepsilon_{\rm y}$ (Fig.~\ref{fig2}g). 

Bcc crystals have much higher friction stresses for dislocation motions than fcc crystals because of their low packing density, high activation energy of vacancy formation, and asymmetric dislocation cores~\cite{hull2011introduction}. We compress fcc-A, fcc-AB$_3$ and bcc-AB single crystals and confirm that $\sigma_{\rm 0}^{\rm bcc-AB}\gg\sigma_{\rm 0}^ {\rm fcc-AB_{3}}>\sigma_{\rm 0}^{\rm fcc-A}$ (Fig.~\ref{fig2}j).  Thus, dislocation density $\rho _\perp^{\rm fcc-A}>\rho _\perp^{\rm fcc-AB_{3}}\gg\rho _\perp^{\rm bcc-AB}$, which is in accordance with Fig.~\ref{fig2}a--f, Supplementary Videos 1--3, and literature~\cite{hull2011introduction}. 

\begin{figure}[h]%
\centering
\includegraphics[width=7.5cm]{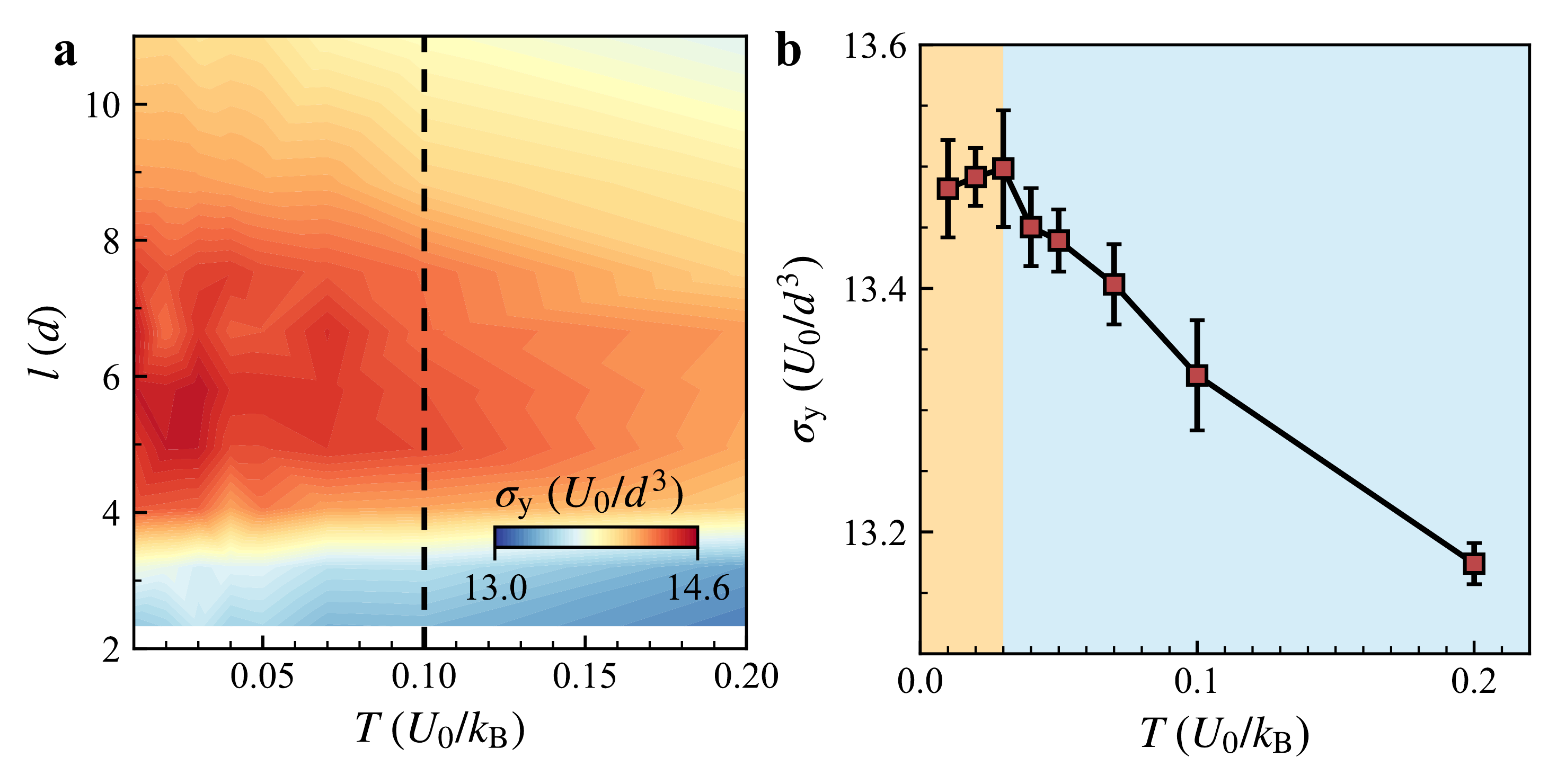} 
\caption{Temperature effect on yield stresses of Type 1 systems. (a) $\sigma_{\rm y}(l,T)$ of samples with $D=45.3$ and different $l$ (i.e. the horizontal dashed line in Fig.~\ref{fig1}c for the HP-IHP boundary). $T$ is well below the melting temperature ($T_{\rm m}\simeq0.5$). $T=0.1$ (dashed line) is used in Figs.~\ref{fig1},~\ref{fig2},~\ref{fig4},~\ref{fig5}. (b) The $\sigma_{\rm y}(T)$ of the polycrystal with $(D,l)=( 45.3, 2.3)$ shows anomalous thermal strengthening (orange) and normal thermal softening (blue) regimes. } \label{fig3}
\end{figure}

The conventional HP and IHP behaviours of $\sigma_{\rm y}(D)$ hold at different temperatures, and polycrystals are soft with low $\sigma_{\rm y}$ at high temperatures~\cite{naik2020hall}. Here, we extend these temperature effects on $\sigma_{\rm y}$ to thick-GB composites for the first time. As temperature $T$ increases, $\sigma_{\rm y}(l)$ still peaks at $l^*=6$ for Type 1 systems (Fig.~\ref{fig3}a) and monotonically decreases for Types 2 and 3 systems (Supplementary Information Fig.~S10). Thus, the generalised HP and IHP behaviours of $\sigma_{\rm y}(D,l)$ hold similarly at different temperatures. The expected thermal softening is observed in the whole temperature range for Type 1 systems with $l>4$ (Fig.~\ref{fig3}a) and all Types 2 and 3 systems (Supplementary Information Fig.~S7). However, the $\sigma_{\rm y}(T)$ of Type 1 systems with $l<4$ exhibits normal thermal softening at $T>0.03$ and anomalous strengthening at $T<0.03$ (Fig.~\ref{fig3}b). This anomalous thermal strengthening was experimentally discovered very recently in pure metals at high strain rates and is attributed to the competition amongst thermal, athermal and drag strengthening mechanisms~\cite{dowding2024metals}. These complex competitions depend on lattice symmetry and composition, which may explain why anomalous thermal strengthening exists in Type 1 polycrystals, but not in other systems.

\begin{figure*}[h]%
\centering
\includegraphics[width=\textwidth]{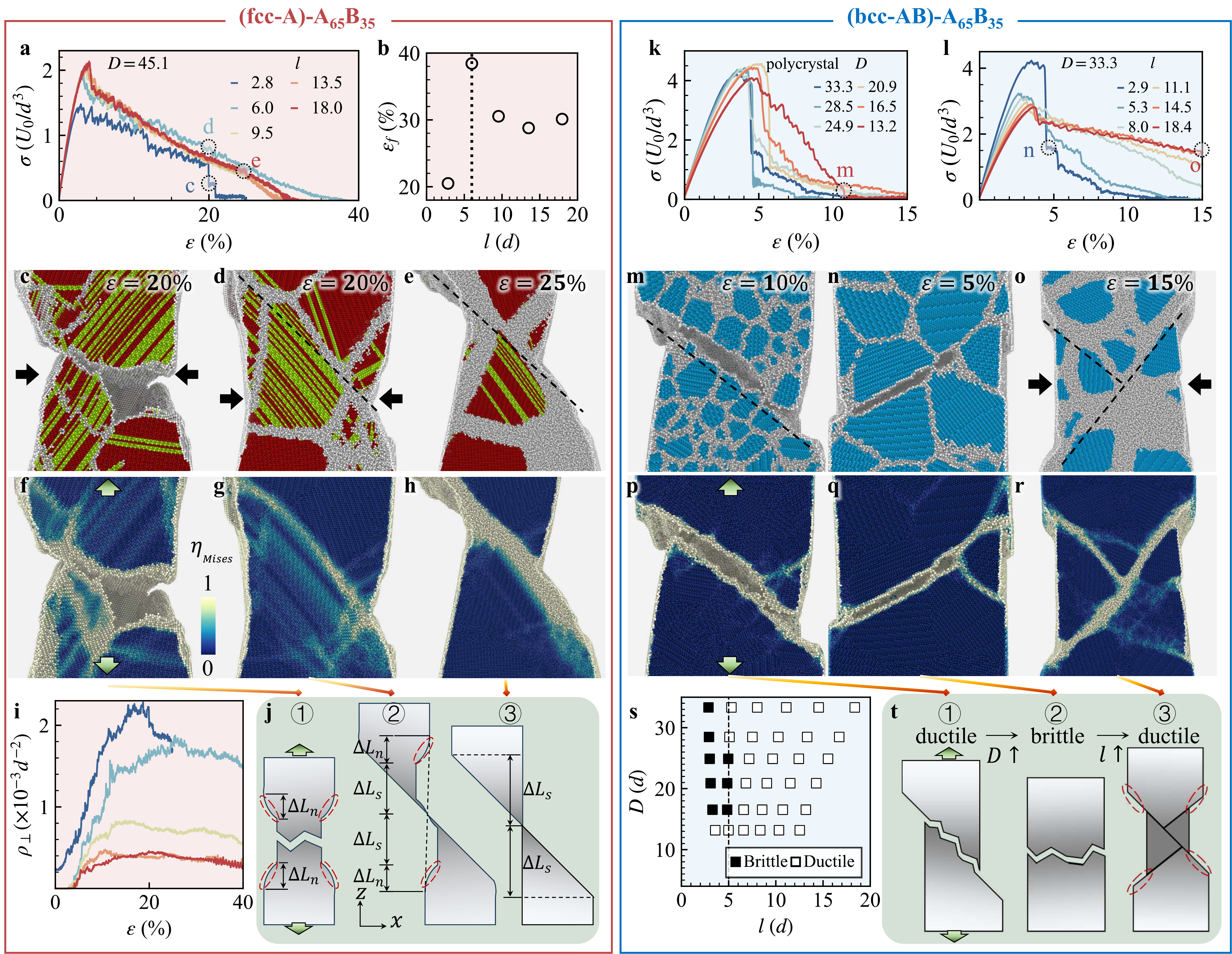}
\caption{Fracture behaviours of Type 1 systems with $D=45.1$ (a--j) and Type 3 systems (k--t). (a) Engineering stress--strain curves $\sigma(\varepsilon)$. (b) Fracture strain $\varepsilon_{\rm f}$ measured at $\sigma=0$ in (a) that reflects ductility. (c-h) Fracture morphologies of three samples with (c, f) $l=2.8$ at $\varepsilon=20\%$, (d, g) $l=6.0$ at $\varepsilon=20\%$ and (e, h) $l=18$ at $\varepsilon=25\%$. Each pair of black arrows in (c, d, o) represents a necking. The black dashed lines in (d, e, m, o) represent shear planes. Each particle is coloured by its local lattice structure (red: fcc; green: hcp; blue: bcc; grey: amorphous) in (c-e) and von Mises shear strain in the corresponding (f--h). (i) Dislocation density in solids with various $l$, seethe legend in (a). (j) Ductility mechanisms sketched by elongation $\Delta L_{\rm n,s}$: \ding{172} necking (highlighted by ellipses) for (c, f), \ding{173} necking followed by shear deformation in GBs for (d, g) and \ding{174} shear deformation in GBs for (e, h). (k) $\sigma(\varepsilon)$ of Type 3 polycrystals with $l\simeq2$. (l) $\sigma(\varepsilon)$ of Type 3 composites with $D=33.3$ and $l>2$. (m--r) Fracture morphologies of three samples with (m, p) $(D,l)=(13.2,3.6)$ at $\varepsilon=10\%$, (n, q) $(D,l)=(33.3,2.9)$ at $\varepsilon=5\%$ and (o, r) $(D,l)=(33.3,18.4)$ at $\varepsilon=15\%$. Particles in (m--o) and (p--r) are coloured in the same way as (c--e) and (f--h), respectively. (s) Ductile-brittle phase diagram of Type 3 systems in $(D,l)$ space. (t) Schematic of ductility mechanisms: \ding{172} single-plane shear deformation followed with a crack for (m, p), \ding{173} crack along GBs for (n, q) and \ding{174} necking induced by chisel edge separation via shear deformation along two perpendicular planes for (o, r). Full images of (c--h) and (m--r) are shown in Supplementary Information Figs.~S13, S14, respectively. The elongation processes of (c--h, m--r) and their $\sigma(\varepsilon)$ are in Supplementary Videos 4--9. \label{fig4}}
\end{figure*}

\section*{Fracture behaviours of composites with different $(D,l)$ }\label{sec3}
Fracture behaviours determine if a solid is ductile or brittle. High ductility is desirable for engineering materials to prevent catastrophic failure during service~\cite{ovid2018review}.  Ductility can be enhanced by introducing twin boundaries, stacking faults, or second-phase precipitates because they can block dislocation motions~\cite{zhu2018ductility} and thus expand the strain hardening regime and delay the onset of fracture~\cite{zhu2004retaining, pineau2016failure1}. However, introducing defects or precipitates usually weakens the material, which is undesirable ~\cite{zhu2004retaining}. Minimising the strength--ductility trade-off is an important challenge in solid mechanics~\cite{ritchie2011conflicts}. Expanding $l$ can enhance ductility because thick GBs act as high-capacity sinks for dislocations and thus expand the strain hardening regime caused by deformations in crystalline grains~\cite{wang2007ductile}. Strength decreases with $l$ in most $(D,l)$ regimes (Fig.~\ref{fig1}c,f,i,l), so the strength-ductility trade-off remains. However, $\sigma_{\rm y}(l)$ increases in $l<l^*$ (Fig.~\ref{fig1}c,l), thus the trade-off is absent in this regime. 

Ductility and the related fracture behaviours of crystalline--amorphous composites have not been studied before. Here, we perform simulations on samples with free surfaces in the $x$ direction and with periodic boundary conditions in the $y$ and $z$ directions under tensile deformation along the $z$ direction. The simulation details are given in Supplementary Information Section 1.2. The abrupt decrease of $\sigma(\varepsilon)$ represents a fracture, that is the appearance of a crack. Note that fracture can also refer to the complete separation of the sample into two or more parts~\cite{knott1973fundamentals}. If no strong plastic deformation occurs before the fracture, the solid is brittle; otherwise, it is ductile~\cite{knott1973fundamentals}.   Ductility can be characterized qualitatively by the fracture morphology and quantitatively by tensile strain $\varepsilon_{\rm f}$ at the fracture~\cite{zhu2018ductility}. 

Type 1 fcc systems are ductile because their stress--strain curves in Fig.~\ref{fig4}a decrease after the peak and reach 0 at large $\varepsilon_{\rm f}$. $\varepsilon_{\rm f}(l)$ in Fig.~\ref{fig4}b peaks at $l=6$. We attribute this maximum ductility at $l=6$ to the fact that the sample has both types of deformation, whereas small- and large-$l$ samples only have one type of deformation, as shown in Fig.~\ref{fig4}j. In small-$l$ samples ($l<6$), such as those in Fig.~\ref{fig4}c,f, the plastic elongation along the $z$ direction is $2\Delta L_{\rm n}$ because of necking (Fig.~\ref{fig4}j\ding{172}). Necking is a type of plastic deformation with a prominent decrease in the local cross-sectional area, as shown in Fig.~\ref{fig4}c and by the ellipses in Fig.~\ref{fig4}j\ding{172}. Necking arises from deformations via considerable dislocation motions and twin-boundary generations in crystalline grains~\cite{knott1973fundamentals}. This result is confirmed by the proliferation of hexagonal close-packed (hcp) lattices (green regions in Fig.~\ref{fig4}c--e) which can be viewed as twin boundaries in fcc lattice. At the end of the necking process, a crack initiates on a free surface then propagates along the shortest GBs that cross the two free surfaces (Supplementary Information Fig.~S13a,b,  Supplementary Video 4). In large-$l$ samples ($l>6$), many dislocations are absorbed into GBs (Fig.~\ref{fig4}i), making the crystalline grains less stretchable. Thus, deformations within GBs pre-empt deformations in crystalline grains (i.e. necking). Plastic deformations mostly occur via shear deformation within thick GBs (Fig.~\ref{fig4}e,h, Supplementary Information Fig.~S13e,f, and Supplementary Video 6), which produce elongation $2\Delta L_{\rm s}$ (Fig.~\ref{fig4}j\ding{174}). The 45$^\circ$ shear plane relative to the tensile elongation direction endures the maximum shear stress ~\cite{greer2013shear}, so the shear deformation is along the GBs whose orientations are close to 45$^\circ$ (dashed lines in Fig.~\ref{fig4}d,e). In the samples with $l=6$, both the necking at the early stage and the 45$^\circ$ shear deformation at the late stage (Fig.~\ref{fig4}d,g, Supplementary Information Fig.~S13c,d and Supplementary Video 5) contribute to the elongation. Thus, the total elongation is $2(\Delta L_{\rm s}+\Delta L_{\rm n})$ (Fig.~\ref{fig4}j\ding{173}), which is greater than those of samples with $l<6$ and $l>6$. This explains the maximum ductility at $l=6$ in Fig.~\ref{fig4}b. Whether the occurrence of the maximum ductility and strength at $l=6$ is a coincidence deserves future studies. 

Unlike the Type 1 fcc samples that are all ductile (Fig.~\ref{fig4}a), the Type 3 bcc samples with different $(D,l)$ can be ductile with large $\varepsilon_{\rm f}$ or brittle with an abrupt decrease in $\sigma(\varepsilon)$ right after the peak (Fig.~\ref{fig4}k,l,s). For brittle samples in Fig.~\ref{fig4}k,l, although $\sigma$ is non-zero after the abrupt drop, a large crack has already formed (Supplementary Video 7), that is, fracture has occurred. The brittle and ductile behaviours of the bcc samples are summarised in Fig.~\ref{fig4}s, with the ductile--brittle transitions occurring at $l\simeq6$ for $D>15$  and at $l<6$ for $D\simeq15$. Ductile--brittle transition occurs in many polycrystals by increasing grain size $D$~\cite{petch1958ductile,li2005ductile,wei2021grain}. Fig.~\ref{fig4}s not only confirms the above conventional $D$ effect, but also shows the $l$ effect on ductile--brittle transition. The fracture morphology is a crack along the GBs with a rough surface for brittle solids (Fig.~\ref{fig4}n,q,t\ding{173}; Supplementary Information Fig.~S14c,d; and Supplementary Video 8) and shear deformation with a rough (Fig.~\ref{fig4}m,p,t\ding{172}; Supplementary Information Fig.~S14a,b; and Supplementary Video 7) or smooth (Fig.~\ref{fig4}o,r,t\ding{174}, Supplementary Information Fig.~S14e,f; and Supplementary Video 9) surface for ductile solids. The crack initiates from a free surface in the ductile samples and from a triple junction inside the bulk in the brittle samples. When the crack on the shear band produces voids (Supplementary Information Fig.~ S14a,b), the subsequent fracture surface becomes rough~\cite{knott1973fundamentals}. A void-free fracture exhibits a smooth surface (e.g. Fig.~\ref{fig4}e,h,o,r). Ductile deformation occurs via a single 45$^\circ$ shear plane along GBs for small-$D$ polycrystals (Fig.~\ref{fig4}m,p,t\ding{172}) and via two shear planes with $\pm45^\circ$ for large-$l$ composites (Fig.~\ref{fig4}o,r,t\ding{174}).
\begin{figure}[h]%
\centering
\includegraphics[width=7.5cm]{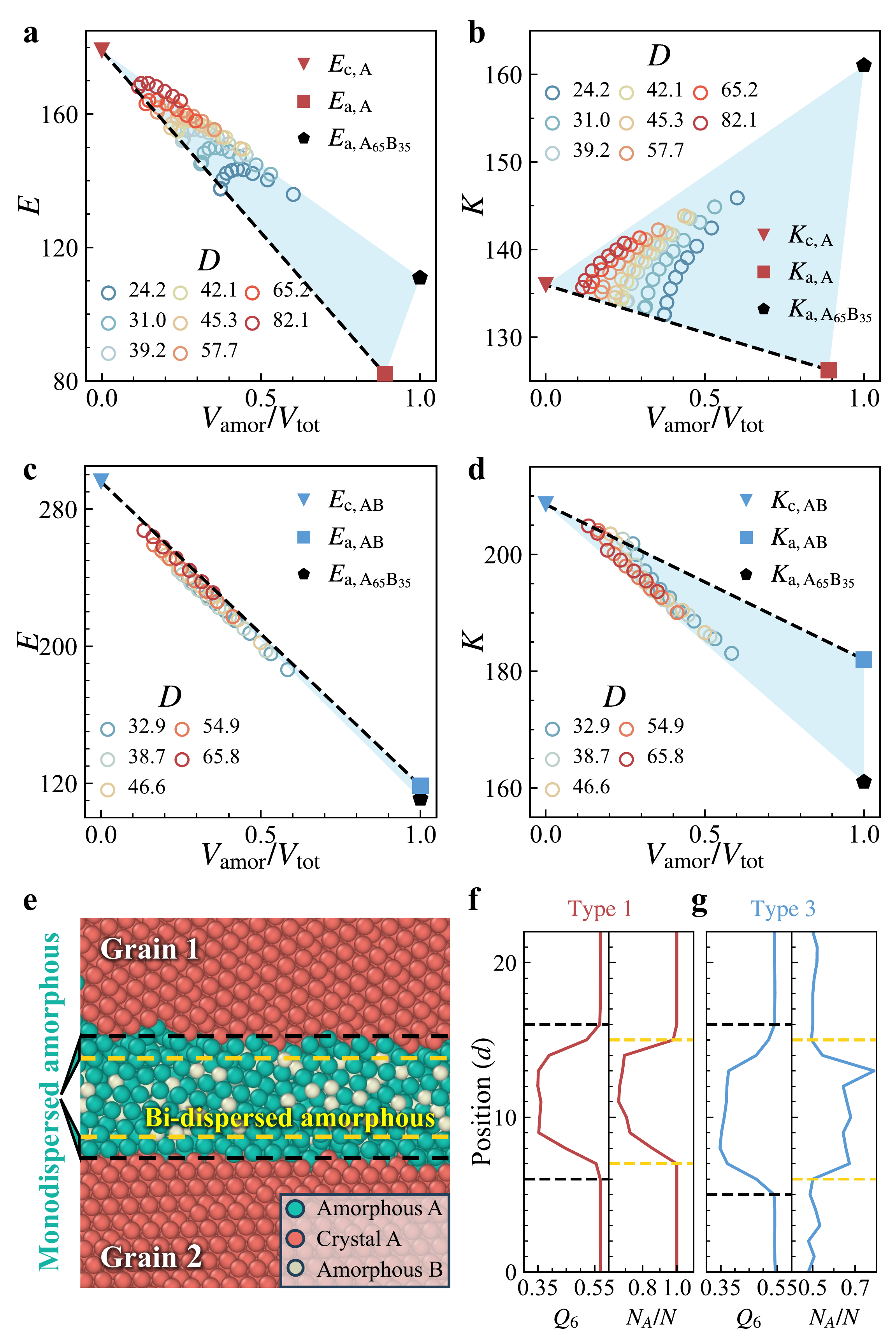}
\caption{Elastic moduli. (a, b) Type 1 systems and (c, d) Type 3 systems. (a, c) Young's modulus $E$ and (b, d) Bulk modulus $K$. All moduli fall into the blue triangular area, which is delimited by the moduli of the single crystal (triangle), purely amorphous glass (square) and interfacial phase with an amorphous structure and crystal composition (pentagon). $V_{\rm amor}/V_{\rm tot}<100\%$ for the red squares in (a, b) because small crystallites inevitably exist in the amorphous sample composed of monodispersed particles. (e) The crystalline--amorphous interface (black dashed lines) and monodispersed--bidispersed interface (yellow dashed lines) are off by one layer of particles, resulting in two monodispersed amorphous interfacial layers. (f, g) Profiles of the mean crystalline order ($Q_6$) and fraction of large particles ($N_{\rm A}/N$) averaged along the interface (f) in Type 1 and (g) Type 3 systems. $Q_6$ and $N_A/N$ rapidly change at slightly different positions (black and yellow dashed lines, respectively).}\label{fig5}
\end{figure}

\section*{Elastic properties in the $(D,l)$ space}\label{sec5}
The elastic properties are also studied in broad $(D,l)$ space. We measure elastic stiffness tensor $C_{ij}$ by using the explicit deformation method~\cite{clavier2017computation} and obtain bulk modulus $K=(C_{11}+2C_{12})/3$, shear moduli $G_1=C_{44}$ and $G_2=(C_{11}-C_{12})/2$ and Young's modulus $E=2C_{44}(C_{11} + 2C_{12})/(C_{11} + C_{12})$. The two shear moduli are very close (Supplementary Information Fig.~S3), indicating that the sample contains a sufficient number of grains to be regarded as isotropic. 

For polycrystals, the elastic modulus is a linear summation of the crystalline and amorphous parts weighted by their volume fractions, i.e., following the simple rule of mixture: $E=\sum_{i=1}^{n}E_{i}f_{i}$~\cite{askeland2003science}, where $E_{i}$ is the elastic modulus and $f_{i}$ is the volume fraction of the $i$th phase. This linear relationship is confirmed in our polycrystal samples: the data point with the lowest amorphous fraction at each $D$, i.e. the polycrystal with the lowest $l$, lies on the dashed line between the triangle (pure crystal) and square (pure amorphous) in Fig.~\ref{fig5}a--d. However, the other data points deviate from the dashed line in Fig.~\ref{fig5}a,b,d. To restore the rule of mixture, we find that the crystalline--amorphous interfaces need to be regarded as the third phase because the interfacial layers have amorphous structures but the same composition as crystalline regions. The crystalline order can be characterized by bond-orientational order parameter $Q_{6}$ (Supplementary Information Eq.~S2)~\cite{steinhardt1983bond}. Fig.~\ref{fig5}f,g shows that $Q_{6}$ and composition change sharply in the normal direction of GB, i.e. sharp crystalline--amorphous and particle composition interfaces. However, the two interfaces are off by one layer of particles, resulting in one layer of the third phase (Fig.~\ref{fig5}f,g). The moduli of the three pure phases measured in separate simulations form a blue triangular area in Fig.~\ref{fig5}a-d that covers all the data points, and each data point is a linear combination of these three phases following the rule of mixture.

\section*{Discussion}\label{sec12}
We measure stress--strain relation $\sigma(\varepsilon)$ (e.g. Fig.~\ref{fig1}b,e,h,k) for four types of samples (Fig.~\ref{fig1}a,d,g,j) with various $(D,l)$ and study the initial elastic regime at small strains (elastic moduli in Fig.~\ref{fig5}), the plastic regime at medium strains (HP and IHP behaviours of $\sigma_{\rm y}$ in Fig.~\ref{fig1}c,f,i,l and their mechanisms in Fig.~\ref{fig2}) and fracture behaviours at large strains (Fig.~\ref{fig4}). These behaviours have been studied as a function of $D$~\cite{pande2009nanomechanics,armstrong201460,cordero2016six,naik2020hall, schuh2007mechanical, ovid2018review,askeland2003science}, but not as a function of $(D,l)$. 

The HP and IHP behaviours are generalised from $\sigma_{\rm y}(D)$ to $\sigma_{\rm y}(D,l)$ (Fig.~\ref{fig1}c,f,i,l). $\sigma_{\rm y}(D)$ under a fixed $l$ is non-monotonic (i.e. with HP and IHP behaviours), with the maximum at a fixed $D=50$ particles for different $l$. $\sigma_{\rm y}(l)$ under a fixed $D$ is non-monotonic for fcc composites and monotonic for bcc composites. Maximum strength $\sigma_{\rm y}^{\rm max}$ is at $(D,l)\simeq (50,6)$ particles for fcc composites with monodispersed crystalline regions, and at $(D,l)\simeq(50,2)$ particles for fcc composites with bidispersed crystalline regions and bcc compositions (Fig.~\ref{fig1}c,f,i,l). These results can explain the recent experimental results on crystalline--amorphous composites. For example, the maximum strength of conventional polycrystals can be exceeded in fcc Ni--Mo crystalline--amorphous composites with 3 nm (i.e. around 10 Mo atoms) GBs~\cite{ding2019thick}, but cannot be exceeded in the bcc Co--Al alloys with 2--10 nm thick GBs~\cite{su2021high}. We suggest that reducing $l$ to 2 nm (i.e. $l^*=6$ atoms) in fcc Ni--Mo composites can further enhance strength.

We find that increasing $l$ and decreasing $D$ have similar effects on $\sigma_{\rm y}$ because they both increase the volume fraction of amorphous structures. However, $\sigma_{\rm y}$ is not solely controlled by the amorphous fraction (Supplementary Information Fig.~S12). The maximum $\sigma_{\rm y}$ is at an amorphous fraction of 28\% in Type 1 systems and 15\% in Types 2--4 systems (Supplementary Information Fig.~S12). At large $D$ and small $l$, plastic deformations primarily arise from dislocation motions inside grains. At small $D$ or large $l$, i.e. high amorphous fraction, plastic deformations mainly occur in GBs via shear deformation (Fig.~\ref{fig2}). Aside from these common features shared by all systems, $\sigma_{\rm y}(l)$ peaks at $l=6$ for fcc composites and monotonically decreases for bcc composites, which is explained as follows. Expanding $l$ up to 6 particles can reduce dislocation motions and thus enhance $\sigma_{\rm y}$ in fcc composites, but it is not effective in bcc composites because dislocations barely move due to their high friction stress (Fig.~\ref{fig2}g). 

The temperature dependence of $\sigma_{\rm y}(l)$ (Fig.~\ref{fig3}) shows that conventional thermal softening in polycrystals also exists in crystalline--amorphous composites, and reveals an abnormal increasing $\sigma_{\rm y}(T)$ regime at low $T$ (Fig.~\ref{fig3}), which was discovered recently in metals~\cite{dowding2024metals}.

For fcc composites, increasing $l$ up to 6 particles enhances strength (Fig.~\ref{fig1}c) and ductility (Fig.~\ref{fig4}b), so the strength--ductility trade-off is avoided. Ductility mainly arises from necking caused by grain deformation in small-$l$ samples and from shear deformation in GBs in large-$l$ samples. Both deformation mechanisms exist in intermediate-$l$ samples, resulting in the maximum ductility at $l=6$ (Fig.~\ref{fig4}b,j). For bcc composites, however, increasing $l$ improves ductility but entails the trade-off of reduced strength. All fcc composites are ductile, and bcc composites exhibit a ductile--brittle transition at $l=6$ for $D>15$ samples and at $l<6$ for $D=15$ samples (Fig.~\ref{fig4}s). 

We find that the rule of mixture for elastic moduli holds well for thick-GB composites. However, when crystalline and amorphous regions have different compositions, their interfaces should be regarded as the third phase to recover the rule of mixture (Fig.~\ref{fig5}). 

The results show that poorly explored structural parameter $l$ has similar importance for mechanical properties as $D$. The results for $(D,l)$ space can guide the fabrication of ultra-strong solids and the optimization of ductility-strength trade-off. Experimentally testing these simulation results is feasible because thick-GB composites can be fabricated in some alloys \cite{gu2024phase, wu2017dual,ding2019thick,su2021high}, glass--ceramics~\cite{holand2019glass} and polymer crystals~\cite{galeski2003strength}. In addition, tuning $(D,l)$ provides a new platform to study ductile--brittle transition and the poorly explored polycrystalline--amorphous transition.

\backmatter

\bmhead{Supplementary information}
Supplementary Information: Sections 1--5, Figs. S1--14, Videos 1--9.

% If your article has accompanying supplementary file/s please state so here. 

% Authors reporting data from electrophoretic gels and blots should supply the full unprocessed scans for key as part of their Supplementary information. This may be requested by the editorial team/s if it is missing.

% Please refer to Journal-level guidance for any specific requirements.

\bmhead{Acknowledgments}

We thank Huijun Zhang for the insightful discussion. This work is supported by RGC grant C6016-20G.

\section*{Declarations}

\begin{itemize}

\item Competing interests. 
The authors declare no competing interests.

\item Availability of data and materials.
The data that support the findings of this study are available from the corresponding author upon reasonable request.
 \item Code availability. 
All simulation codes are available from the corresponding author upon reasonable request. 

\item Authors' contributions.
Y.H. and Z.X. conceived and designed the research. Z.X. performed the simulations and data analysis with help from Y.H. and M.L. Z.X. and Y.H. wrote the paper. Y.H. supervised the work. All authors discussed the results.

\end{itemize}

%\bibliography{sn-bibliography}
%\bibliography{extractred_bib_SI}
%\bibliographystyle{naturemag}

\end{document}

\doublespacing
\maketitle
\newpage
\setlength{\cftsecindent}{0.5cm}
\setlength{\cftsubsecindent}{1.5cm}
% \tableofcontents
\singlespacing
\begin{spacing}{2.0}
    \tableofcontents
    % \listoffigures
\end{spacing}

\newpage

\section{Method}
\subsection{Construction of polycrystals with thick grain boundaries (GBs)}
\begin{figure}[h]
\centering
\includegraphics[width=\textwidth]{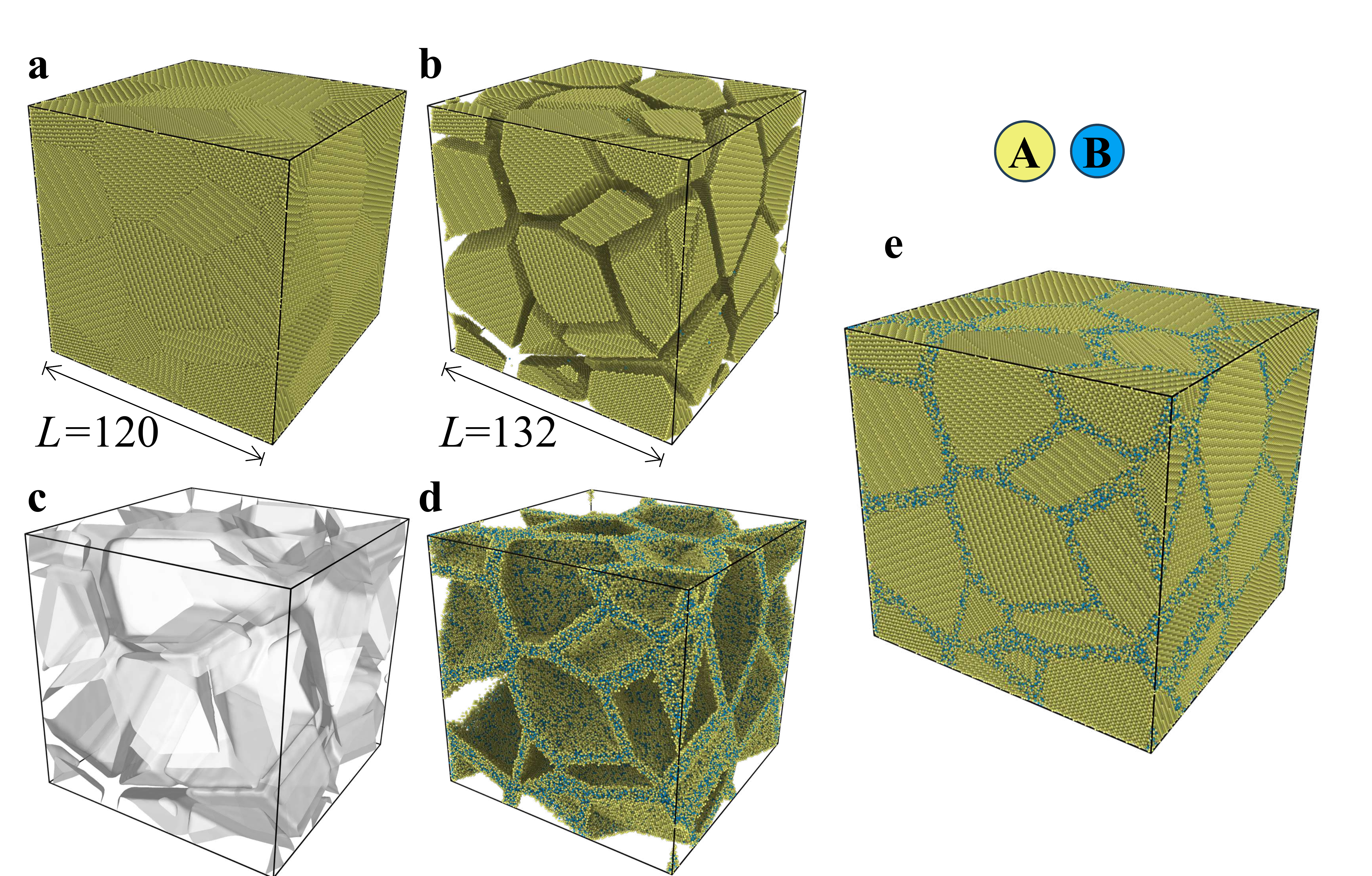}
\caption{\label{SI1} Procedure of constructing a crystalline--amorphous composite from a polycrystal. (a) Initial binary polycrystal with mean grain diameter $D=46.57$ in a cubic box with side length $L=120$. (b) Shifted grains in the expanded cubic box with $L=132$. (c) Crystalline--amorphous interfaces in (b). (d) In an amorphous solid composed of A$_{65}$B$_{35}$ mixture of Lennard--Jones particles, the particles in the region given by (c) are chosen. (e) Final sample with thick GBs obtained by combining (b) and (d).}
\end{figure}

In previous simulations, crystalline--amorphous composites were constructed by quenching liquids~\cite{brink2018metallic}, substituting surface layers of crystalline grains to amorphous structures~\cite{xiao2021mitigating} or using periodic grain patterns~\cite{qian2023amorphous}. These methods do not produce a series of samples with the same crystalline regions and different GB thicknesses, or the grains are too regular to mimic real materials. Here, we propose a method to create the desired GB thickness and maintain the same sizes and irregular shapes of crystalline grains so that the GB thickness effect can be well measured by comparing a series of samples. As illustrated in Fig.~\ref{SI1}, the crystalline--amorphous composite is constructed through the following steps. Firstly, a polycrystal is generated using the Voronoi tessellation method~\cite{hirel2015atomsk} (Fig.~\ref{SI1}a). The shape of each grain is given by the Voronoi cell. The lattice orientations of the grains are set as random. Secondly, the simulation box is expanded, and the centre of mass of each grain is proportionally projected into the new box. Such a shift of grains creates gaps with uniform thickness $l$ between grains (Fig.~\ref{SI1}b). Thirdly, surface meshes are created by $\alpha$-shape method~\cite{edelsbrunner1994three} to describe the shape of gaps (Fig.~\ref{SI1}c). Fourthly, the surface meshes are used to extract thick-GB regions from an amorphous glass (Fig.~\ref{SI1}d). This glass is created by rapidly cooling a liquid from $3.5T^*$ to $0.1T^*$ within $100\tau$ under $10P^*$. Last, the displaced grains in Fig.~\ref{SI1}b are combined with the GBs in Fig.~\ref{SI1}d to produce a thick-GB sample (Fig.~\ref{SI1}e). The number of particles ranges from 2.5  million to 11.4 million for different samples. The structures in Figs.~1a,d,g,j,~2a--f,~4c--h,m--r of the main text and Figs.~\ref{SI1},~\ref{SI10}a,~\ref{SI13},~\ref{SI14} are visualised using the OVITO visualisation tool~\cite{stukowski2009visualization}.

\begin{figure}[t]
\centering
\includegraphics[width=\textwidth]{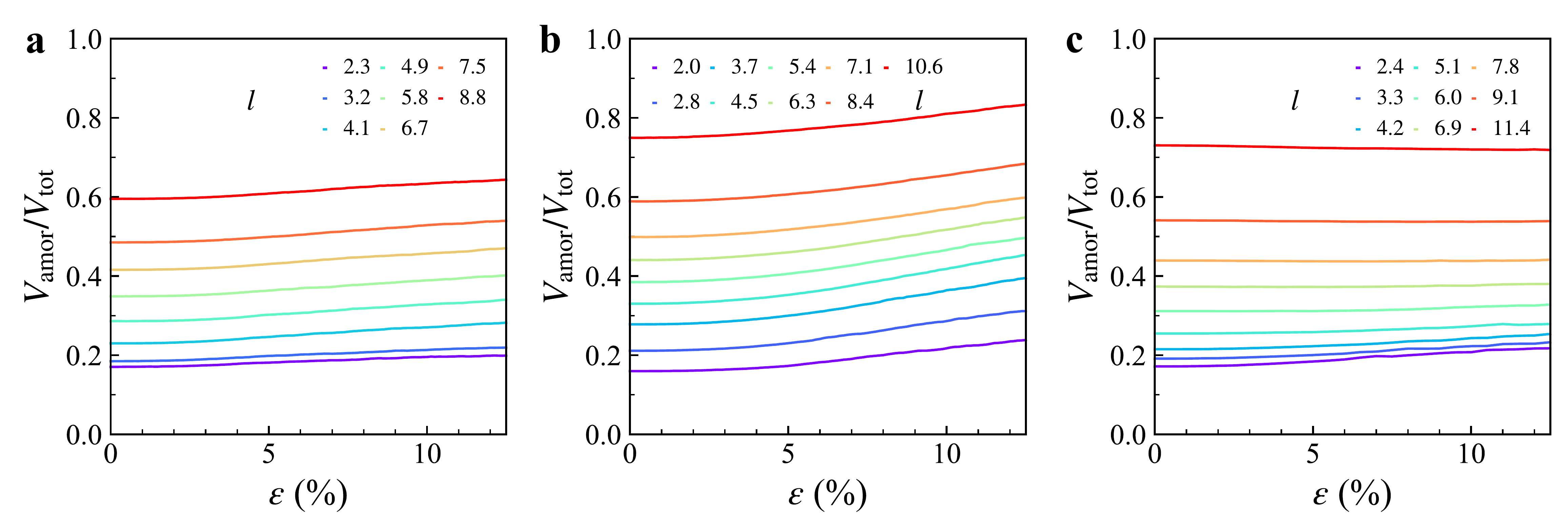}
\caption{\label{SI2} Fraction of the amorphous phase under loading strain $\varepsilon$. (a) Type 1 systems with $D=45.3$. (b) Type 2 systems with $D=46.9$.  (c) Type 3 systems with $D=46.6$.}
\end{figure}

\subsection{Simulation}
 The initial polycrystals are in the same-sized cubic box ($L=120d$), so a polycrystal with large-$D$ grains contains a small number of grains (Fig.~\ref{SI1}). The diameter of the large particle, $d$, is used as the length unit. After GBs with different thicknesses are inserted, the box sizes become slightly different. $(D,l)$ are tuned in four types of systems (Fig.~1a,d,g,j of the main text). (1) The face-center-cubic (fcc) crystalline grains composed of large (A-type) particles and amorphous GBs composed of large and small (B-type) particles with a number ratio of $N_{\rm A}:N_{\rm B}=65:35$, i.e. (fcc-A)-(A$_{65}$B$_{35}$). The number ratio $N_{\rm A}:N_{\rm B}=65:35$ can form stable amorphous structures without phase separation~\cite{bruning2008glass}, so it is used in all amorphous regions. (2) The second type is a binary solid with $N_{\rm A}:N_{\rm B}=1:3$ in fcc crystalline grains and $ N_{\rm A}: N_{\rm B}=65:35$ in amorphous GBs, i.e. (fcc-AB$_{3}$)-(A$_{65}$B$_{35}$). (3) The third type is a binary solid with $N_{\rm A}:N_{\rm B}=1:1$ in the body-centered-cubic (bcc) crystalline grains and $ N_{\rm A}: N_{\rm B}=65:35$ in amorphous GBs, i.e., (bcc-AB)-(A$_{65}$B$_{35}$). (4) The fourth type refers to fcc crystalline grains composed of copper (Cu) atoms and amorphous GBs composed of a copper--zirconium (Cu--Zr) mixture with $N_{\rm Cu}:N_{\rm Zr}=64:36$, i.e. (fcc-Cu)-(Cu$_{64}$Zr$_{36}$). 

All simulations are performed with the Large-scale Atomic/Molecular Massively Parallel Simulator (LAMMPS)~\cite{plimpton1995fast}. Types 1--3 systems are composed of particles with the Lennard--Jones (LJ) potential. The interactions between binary particles are set to follow the Kob--Anderson mixture~\cite{kob1995testing}: $d_{\rm AA}=1.0$, $d_{\rm BB}/d_{\rm AA}=0.88$ and $d_{\rm AB}/d_{\rm AA}=0.8$; $U_{\rm AA}=1.0$, $U_{\rm BB}/U_{\rm AA}=0.5$ and $U_{\rm AB}/U_{\rm AA}=1.5$; all masses are set to $m=1.0$. Pressure, time and temperature have units $P^*= U_0/d^3$, $\tau=\sqrt{{md^3}/{U_0}}$ and $T^*=U_0/k_{\rm B}$, respectively. Boltzmann constant $k_{\rm B}=1$. The Types 1--3 samples are well annealed for $700\tau$ at $0.1T^*$ and $10P^*$. The mean-square displacement and structures of types 1-3 systems do not change after $600\tau$ equilibration, indicating that the samples are well relaxed. Type 4 systems are composed of Cu and Zr atoms with multibody embedded-atom-method potentials~\cite{mendelev2019development}. Type 4 samples are relaxed at $T=300$~K and $P=0$ for 1~ns. Uniaxial compression is applied on Types 1--3 LJ samples with strain rate $\dot{\varepsilon}=2.5\times10^{-4}{t}^{-1}$ and Type 4 Cu--Zr samples with $\dot{\varepsilon}=2.5\times10^{8}{\rm s}^{-1}$.

\subsection{Sample characterisations}

Common neighbour analysis (CNA)~\cite{honeycutt1987molecular} is implemented to identify each particle's local lattice symmetry or defect type. Crystalline regions are barely amorphised during the deformation process (Fig.~\ref{SI2}). Thick GBs occasionally contain a few tiny ($D<3$) crystallites that are not counted as grains. Mean grain diameter $D\equiv(\sum_i {\sqrt[3]{6V_{i}/\pi}})/N_i$, where $V_i$ and $N_i$ are the volume and number of particles of grain $i$, respectively. GB thickness $l\equiv2V_{\rm GB}/A{_{\rm c}}$, where $V_{\rm GB}$ is the total volume of the thick amorphous GBs and $A_{\rm c}$ is the total surface area of all crystalline grains. $V_{\rm GB}$ and $A_{\rm c}$ are calculated by the surface geometry construction method (Fig.~\ref{SI1})~\cite{stukowski2014computational}. The presence of triple junctions increases the mean GB thickness, so the measured $l\simeq2$ instead of 1 in polycrystals. 

\begin{figure}[h]
\centering
\includegraphics[width=\textwidth]{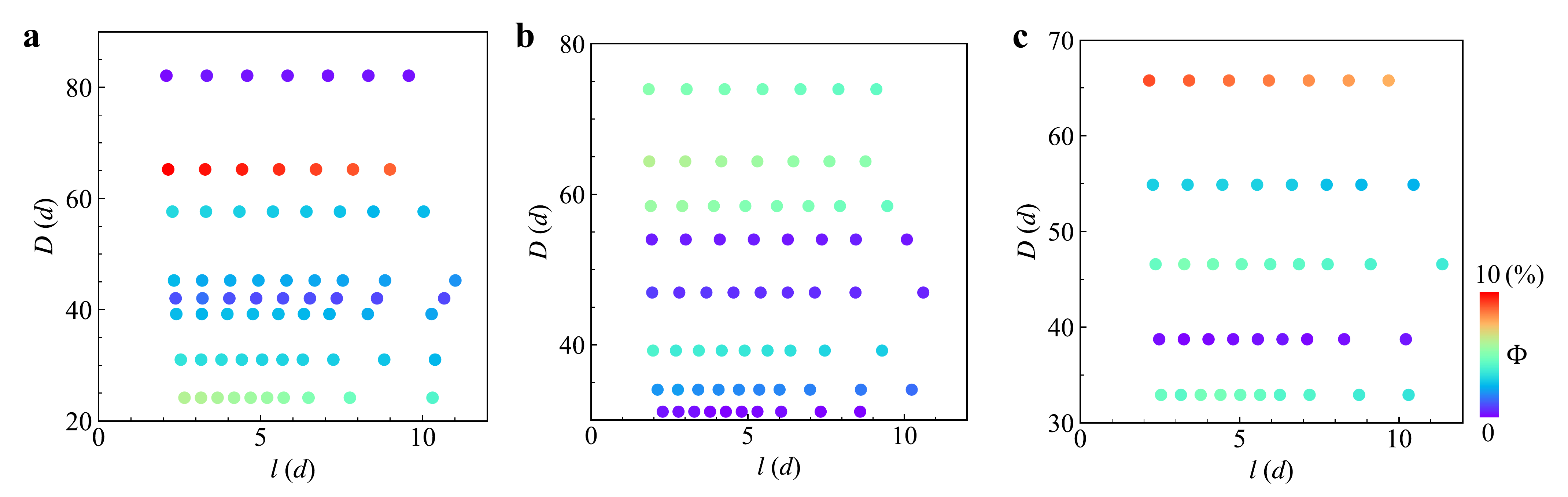}
\caption{\label{SI3} Degree of sample anisotropy $\Phi$ for (a) Type 1, (b) Type 2 and (c) Type 3 systems. $\Phi=0$ means that the sample is fully isotropic.}
\end{figure}

Von Mises shear strain is commonly used to characterise local shear strain~\cite{mises1928mechanik}. To visualise the deformation mechanism, we measure the von Mises shear strain as follow:
\begin{equation}
\eta_{\mathrm{Mises}}=\sqrt{\eta_{\mathrm{xy}}^2 + \eta_{\mathrm{yz}}^2 + \eta_{\mathrm{zx}}^2 + \frac{(\eta_{\mathrm{xx}} - \eta_{\mathrm{yy}})^2 + (\eta_{\mathrm{yy}} - \eta_{\mathrm{zz}})^2 + (\eta_{\mathrm{zz}} - \eta_{\mathrm{xx}})^2}{6}}.
\end{equation}
$\eta_{ij}$ is the Green-Lagrangian strain tensor calculated by $\eta_{ij} = (\mathbf{F}^{\mathrm{T}} \mathbf{F} - \mathbf{I})/2$~\cite{lubliner2008plasticity}, where $\mathbf{F}$ is the deformation gradient tensor.  $\mathbf{I}$ is the $3\times3$ identity tensor, with ones on the diagonal and zeros elsewhere. 

The fracture behaviours are investigated in separate simulations of Type 1 fcc and Type 3 bcc systems. The sample in a box with a side length ratio of $x:y:z=1:1:2.4$ is relaxed under $0.5T^{*}$ for 400$\tau$ and under $0.005T^{*}$ for another 400$\tau$ at $P=0$ in the NPT ensemble. After relaxation, tensile deformation is applied along the $z$ direction with a constant strain rate $\dot{\varepsilon}=1\times10^{-5}\tau^{-1}$ for at least 3000$\tau$ until a fracture forms. The small distance between the two free surfaces ($\simeq100$ particles) results in a pronounced surface effect \cite{gao1999mechanism}, which is a common problem in simulation studies on fracture behaviours due to the finite system size~\cite{brink2018metallic}.

We use the explicit deformation method to measure elastic stiffness tensor $C_{ij}$ because it has better accuracy~\cite{clavier2017computation} than the strain-fluctuation~\cite{parrinello1982strain} and stress-fluctuation methods~\cite{lutsko1989generalized}. On the basis of $C_{ij}$, we calculate bulk modulus $K=(C_{11}+2C_{12})/3$ and Young's modulus $E=2C_{44}(C_{11} + 2C_{12})/(C_{11} + C_{12})$ (Fig.~5 in the main text).

The local crystalline order of particle $i$ is characterised by its bond-orientational order parameter: 
\begin{equation}
Q_{i, \iota}=\sqrt{\frac{4\pi}{2\iota+1}\sum_{m=-\iota}^{\iota}\left|\sum_{j=1}^{n}\frac{A_{j}}{A}Y_{\iota m}\left(\theta_{j},\phi_{j}\right)\right|^{2}},
\end{equation}
where $\theta_{j}$ and $\phi_{j}$ are spherical polar angles of the bond vector from particle $i$ to its $j$th neighbour, $n$ is the number of nearest neighbours of particle $i$, $A_j$ is the area of the Voronoi facet to the $j$th neighbour. $A$ is the total surface area of the Voronoi cell and $Y_{\iota m}$ is a spherical harmonic function of degree $\iota$ and order $m$. We use sixfold ($\iota=6$) bond-orientational order parameter $Q_{6}$ to distinguish crystalline and amorphous phases.

When grains are large, the sample contains a small number of grains, which makes the sample not fully isotropic. The degree of anisotropy is characterised by the modified Zener ratio~\cite{li1987single} $\Phi=|G_{1}-G_{2}|/(G_{1}+G_{2})$, where $G_{1}=(C_{11}-C_{12})/2$ and $G_{2}=C_{44}$ are the two shear moduli. $\Phi_{\rm iso}<5\%$ for most samples, and the maximum is about 10$\%$ (Fig.~\ref{SI3}). Thus the samples are quite isotropic. Moreover, each data point of $\Phi_{\rm iso}$ in Fig.~\ref{SI3} is from a single measurement, and we minimise the effect of anisotropy by averaging the results of uniaxial compression in $x$, $y$ and $z$ directions.
\newpage
\section{Generalised HP and IHP behaviours}
\subsection{Amorphous fractions on the contour plots of $\sigma_{\rm y}(D,l)$}
\begin{figure}[h]
\centering
\includegraphics[width=14cm]{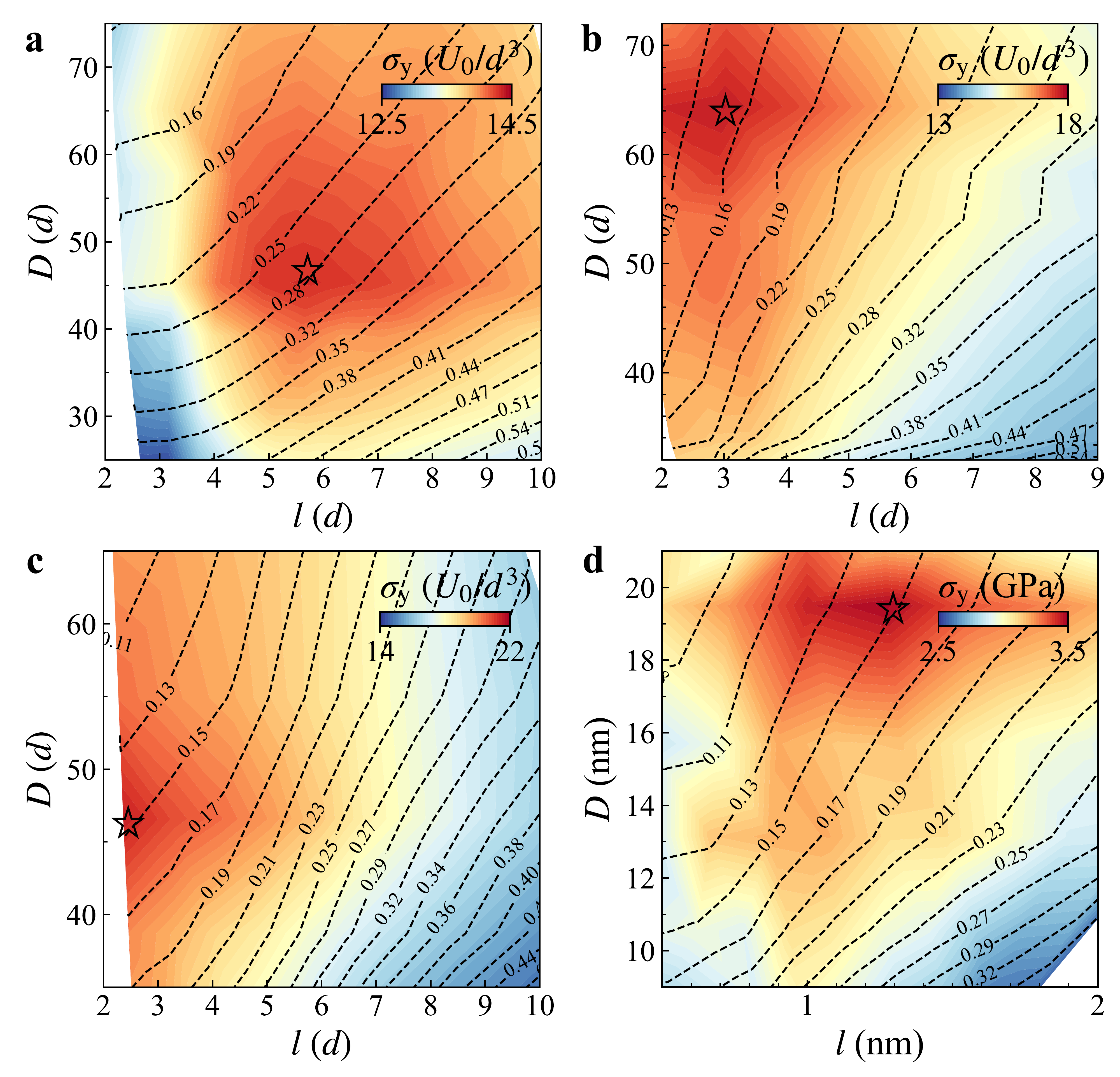}
\caption{\label{SI4} Contour maps of amorphous volume fraction $\phi_{\rm amor}$ (black curves) overlaid on contour maps of $\sigma_{\rm y}$ for systems 1--4 in Fig.~1c,f,i,l of the main text. $\sigma_{\rm y}^{\rm max}$ labelled with $\medwhitestar$ is at $\phi_{\rm amor}=28\%$ in Type 1 systems and at $\phi_{\rm amor}=15\%$ in Types 2--4 systems.}
\end{figure}
The ratio of the volume of amorphous regions to the volume of the whole sample, $\phi_{\rm amor}$, is labelled on $\sigma_{\rm y}(D,l)$ shown in Fig.~S4. When $\phi_{\rm amor}$ is fixed, $\sigma_{\rm y}(l)$ and $\sigma_{\rm y}(D)$  are usually non-monotonic, and the maximum $\sigma_{\rm y}$ is achieved at different $(D,l)$ under different $\phi_{\rm amor}$ (Fig.~S4).
\subsection{Using flow stress as strength}
\begin{figure}[h]
\centering
\includegraphics[width=14cm]{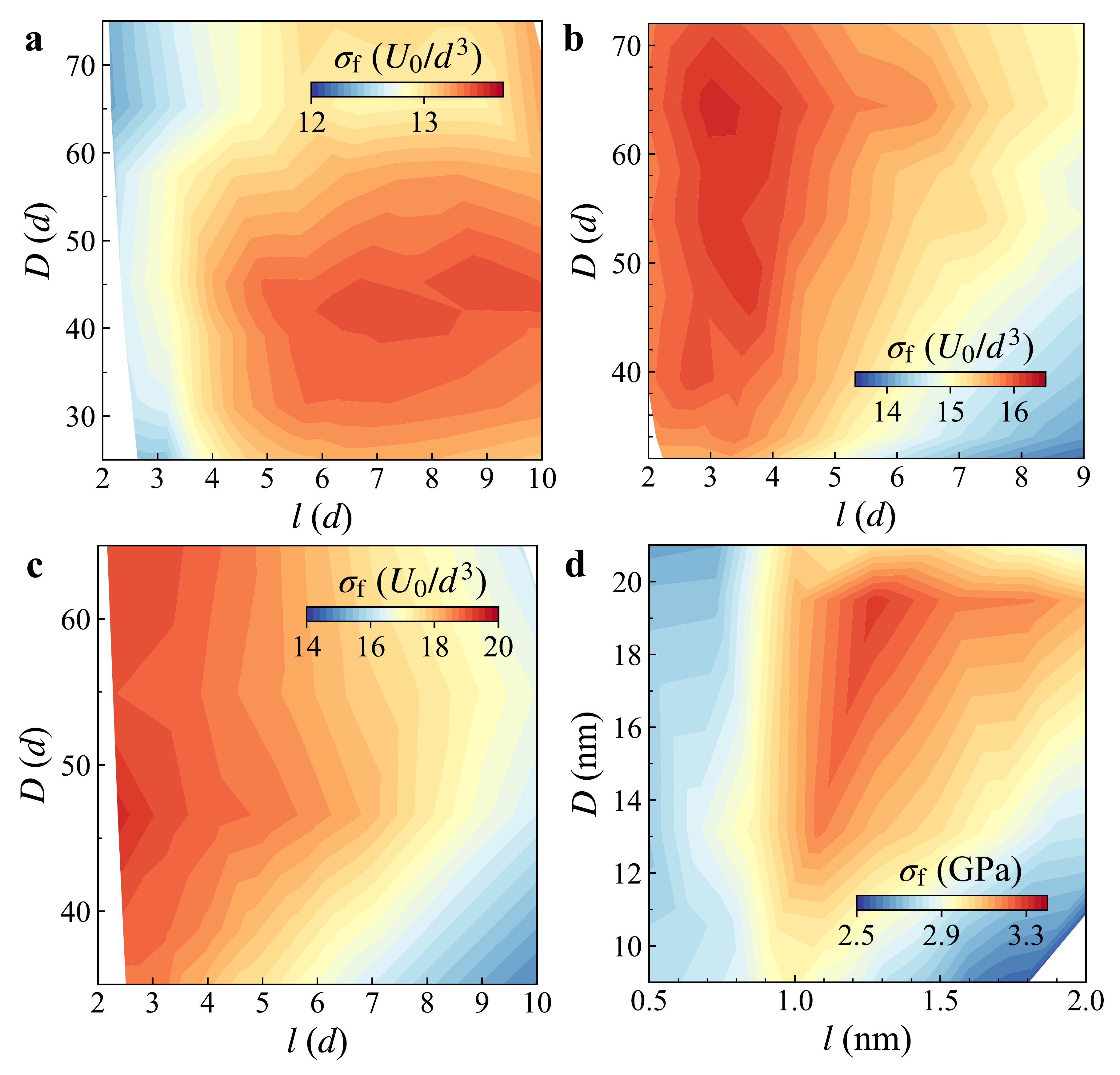}
\caption{\label{SI5} Contour maps of flow stress $\sigma_{\rm f}(D,l)$ for samples with different mean grain diameters $D$ and GB thicknesses $l$. $d$ is the diameter of large particles. (a) Type 1 (fcc A)-(A$_{65}$B$_{35}$) systems. (b) Type 2 (fcc-AB$_{3}$)-(A$_{65}$B$_{35}$) systems. (c) Type 3 (bcc-AB)-(A$_{65}$B$_{35}$) systems. (d) Type 4 (fcc-Cu)-(Cu$_{64}$Zr$_{36}$) systems. The corresponding contour maps of yield stress $\sigma_{\rm y}(D,l)$ are in Fig.~1c,f,i,l of the main text.}
\end{figure}

The strength of a solid can be defined as the maximum stress (i.e. yield stress $\sigma_{\rm y}$)~\cite{brink2018metallic} or flow stress~\cite{schiotz2003maximum}. Both definitions are commonly used in HP/IHP behaviours. The stress--strain curves of some samples decrease continuously without a plateau (e.g. Figs.~1a,b,c, 4a,k of the main text), so flow stress is not rigorously defined. Nevertheless, we define $\sigma_{\rm f}$ as the average stress in $0.1<\varepsilon<0.125$ for Types 1--3 systems and as the average stress in $0.08<\varepsilon<0.1$ for Type 4 systems (i.e. the plateau regimes in Fig.~1b,e,h,k of the main text). $\sigma_{\rm y}$ and $\sigma_{\rm f}$ are measured by averaging the results of the uniaxial compressions along $x$, $y$ and $z$ directions in the samples with different initial thermal motions of particles for sufficient statistics. Their contour plots are shown in Fig.~1c,f,i,l of the main text for $\sigma_{\rm y}$ and in Fig.~\ref{SI5} for $\sigma_{\rm f}$.

\subsection{Effects of strain rate and relaxation time on $\sigma_{\rm y}(D,l)$}
\begin{figure}[h]
\centering
\includegraphics[width=\textwidth]{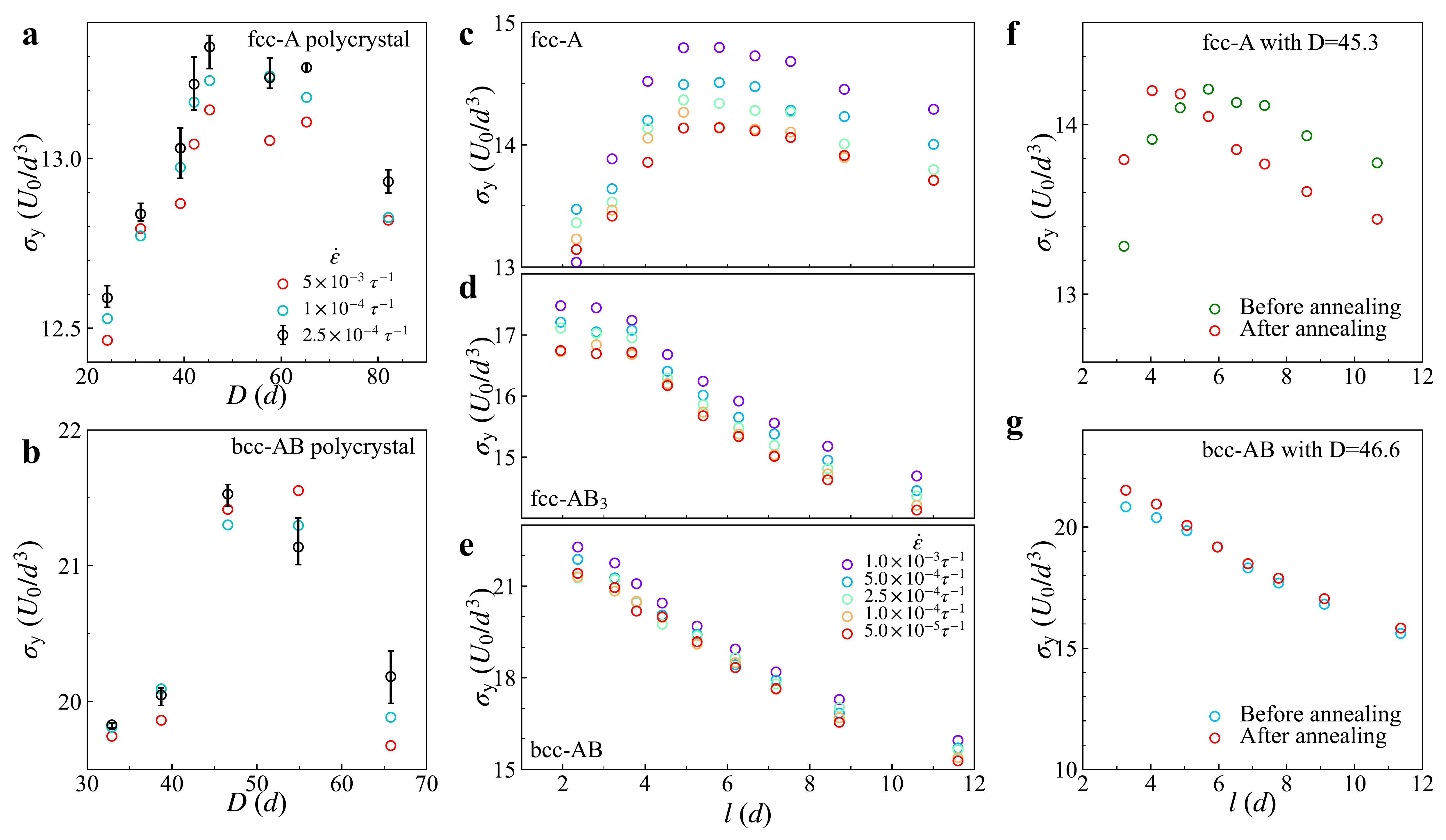}
\caption{\label{SI6} (a, b) Yield stress $\varepsilon_{\rm y}$ under different strain rates $\dot{\varepsilon}$ and mean grain diameters $D$. (a) fcc-A and (b) bcc-AB polycrystals ($l\simeq 2$). (a, b) share the same legend. (c-e) Yield stress under different strain rates for (c) Type 1 fcc-A systems with $D=45.3$, (d) Type 2 fcc-AB$_3$ systems with $D=46.9$ and (e) Type 3 bcc-AB systems with $D=46.6$. (c, d, e) share the same legend. (f, g) Yield stress of (f) fcc-A and (g) bcc-AB polycrystals before and after annealing under $T^*$ for $3000\tau$.}
\end{figure}

Our strain rate $\dot{\varepsilon}$ is comparable to those used in all recent simulations about the deformation of crystal--amorphous composites (e.g., refs.~\cite{xiao2021mitigating,qian2023amorphous}) and a few laser-induced impact experiments~\cite{dowding2024metals,lee2012high}. However, $\dot{\varepsilon}$ is much higher than that in real experiments~\cite{gu2024phase} because low strain rates in real experiments are computationally too expensive. Excessive deformation rates may affect material strength and the deformation mechanism~\cite{fan2012onset}. Nevertheless, simulations at strain rates have been well accepted as an effective method for the study of conventional HP and IHP behaviours (e.g., refs.~\cite{schiotz2003maximum,wolf2005deformation,bringa2005ultrahigh, zepeda2017probing,cao2022maximum}) because they can reproduce the experimental HP--IHP boundary ($D^*$). Different strain rates shift the $\varepsilon_{\rm y}(D)$ curves but do not substantially affect the peak position, i.e. the HP--IHP boundary~\cite{carlton2007behind}. Our results confirm that the peak position of $\varepsilon_{\rm y}(D)$ is robust under different strain rates (Fig.~\ref{SI6}a,b). Similarly, the peak position of $\varepsilon_{\rm y}(l)$ is robust under different strain rates (Fig.~\ref{SI6}c,d,e). 

Previous simulation studies have demonstrated that yield stress is influenced by relaxation time, temperature and annealing time~\cite{valiev2000bulk}. We compare the yield stresses of samples with and without annealing in Fig.~\ref{SI6}f,g. Annealing shifts the peak position of $\sigma_{\rm y}(l)$ towards a low $l$ for fcc samples (Fig.~\ref{SI6}f). We attribute it to the minimal dislocations after annealing, which produces a narrow dislocation-motion-dominated HP-like regime. Annealing has no substantial effect on bcc samples (Fig.~\ref{SI6}g) because dislocations rarely move in bcc crystals~\cite{hull2011introduction}.

\subsection{Temperature effect on $\sigma_{\rm y}(D,l)$}
\begin{figure*}[h]%
\centering
\includegraphics[width=10cm]{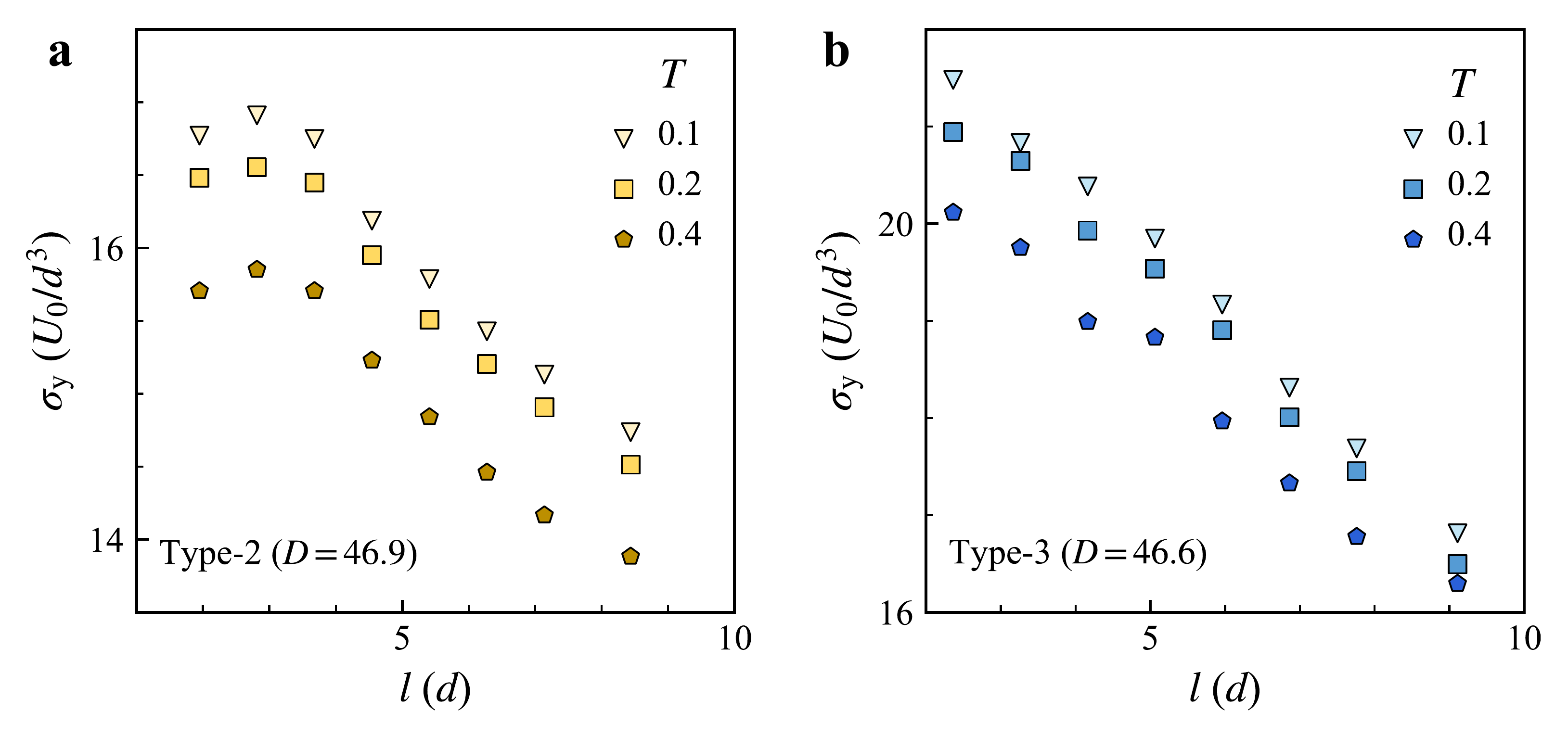} 
\caption{\label{SI7}Temperature effect on $\sigma_{\rm y}(l)$ for (a) Type 2 systems with $D=46.9$ and (b) Type 3 systems with $D=46.6$.} 
\end{figure*}
\newpage
\section{Mechanisms of the generalised HP and IHP behaviors}
\subsection{Ratio of deformations in GBs and in crystalline grains}

\begin{figure}[h]
\centering
\includegraphics[width=6cm]{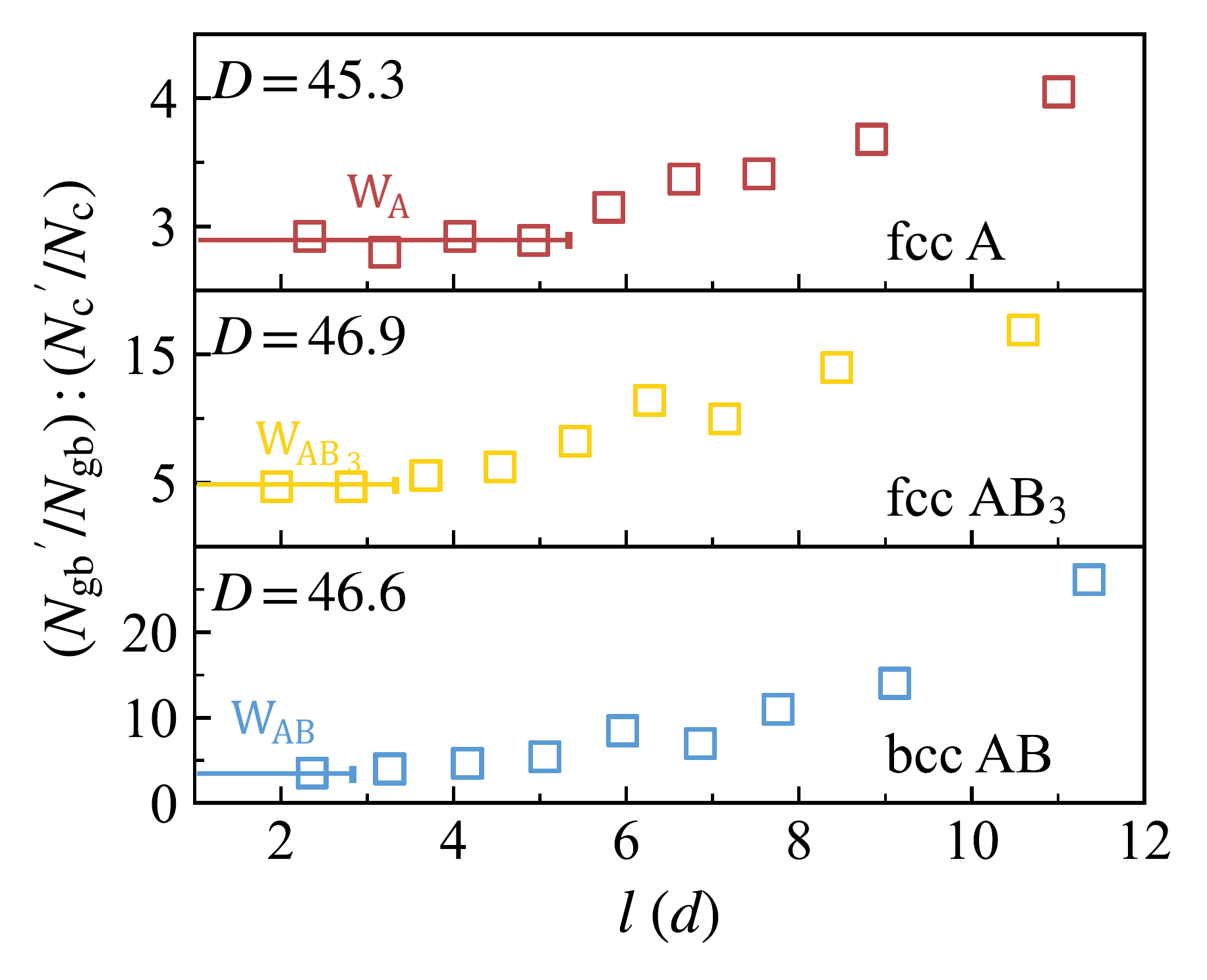}
\caption{\label{SI8} Ratio of the deformation intensities of the amorphous and crystalline regions. The top to bottom panels are for types 1-3 systems. $N'$ is the number of particles with $\eta_{\rm Mises}>0.12$ amongst the $N$ particles in GBs (subscript gb) or crystals (subscript c). The horizontal lines denote the plateau widths $W_{\rm A}$, $W_{\rm AB_3}$ and $W_{\rm AB}$.}
\end{figure}
Dislocation motions in crystalline grains dominate the deformation in the conventional HP regime, whereas GB deformations dominate in the conventional IHP regime~\cite{pande2009nanomechanics}. To compare the two types of deformations in the generalised HP/IHP behaviours of $\sigma_{\rm y}(D,l)$, we measure the number ratio of particles that carry dislocation motions $N_{\rm c}'$ to that in all crystalline grains $N_{\rm c}$ and the number ratio of particles that carry GB deformations $N_{\rm gb}'$ to that in all amorphous GBs $N_{\rm gb}$. The ratio between the two ratios at $\sigma_{\rm y}^{\rm max}(l)$ (horizontal dashed lines in Fig.~1c,f,i of the main text) is shown in Fig.~\ref{SI8}. For Type 1 systems, the ratio of plastic deformation from GBs to that from crystalline regions is constant at $l<W_{\rm A}=6$ and increases at $l>6$ (Fig.~\ref{SI8}). The two regimes coincide with the dislocation-motion-dominated HP-like regime and shear-transformation-zone or shear-band-dominated IHP-like regimes~\cite{qian2023amorphous} of $\sigma_{\rm y}(l)$ at a fixed $D$ in Fig.~1c of the main text. Shear-transformation zones are localized atomic or molecular deformation patches induced by shear, and the shear bands are large zones of intense shear strain. The dislocation-dominated-deformation regime is at $l< W_{\rm B}=3$ for Type 2 systems (Fig.~\ref{SI8}), which is narrower than the $l<6$ regime for type 1 systems because Type 2 systems have less dislocations (Fig.~2d,e of the main text). Type 3 bcc systems are nearly dislocation free (Fig.~2g of the main text), so GB deformation dominates and $(N_{\rm gb}'/N_{\rm gb}):(N_{\rm c}'/N_{\rm c})$ in Fig.~\ref{SI8} monotonically increases with $l$. The plateau regimes at $l<W=6,3,2$ particles in Fig.~\ref{SI8} correspond well to $l^*\simeq6,3,2$ at $\sigma_{\rm y}^{\rm max}$ ($\medwhitestar$ in Fig.~1c,f,i of the main text). This result confirms the dislocation-motion-dominated and GB-dominated deformation mechanisms in the increasing and decreasing regimes of $\sigma_{\rm y}^{\rm max}(l)$, respectively. In summary, the plateau width $W_{\rm fcc-A}>W_{\rm fcc-AB_3}>W_{\rm bcc-AB}$ (Fig.~\ref{SI8}) is in accordance with $\rho_{\perp}^{\rm fcc-A}>\rho_{\perp}^{\rm fcc-AB_3}>\rho_{\perp}^{\rm bcc-AB}$ in Fig.~2d-f,h of the main text and explains $l^*_{\rm fcc-A}>l^*_{\rm fcc-AB_{3}}>l^*_{\rm bcc-AB}$ in Fig.~1c,f,i of the main text.  

\begin{figure}[t]
\centering
\includegraphics[width=\textwidth]{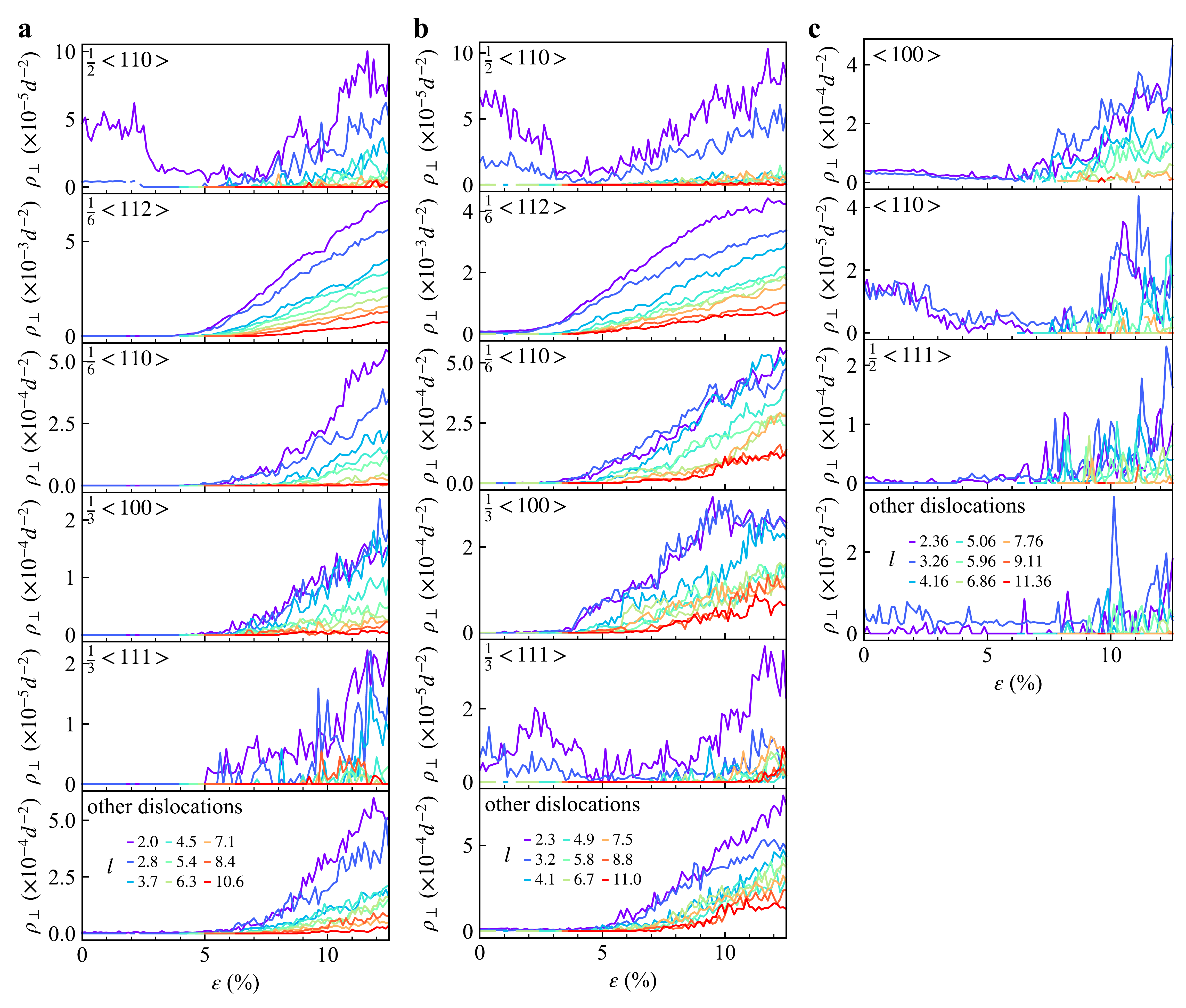}
\caption{\label{SI9} Densities of various types of dislocations under uniaxial strain $\varepsilon$. (a) Type 1 fcc systems with $D=45.3$. (b) Type 2 fcc systems with $D=46.9$.  (c) Type 3 bcc systems with $D=46.6$. Partial dislocations (e.g. $\frac{1}{6}\langle112\rangle$) rarely exist in bcc lattices, so they are not shown in (c). The legends are in the bottom panels.}
\end{figure}

\subsection{Dislocation density during deformation}

Different types of dislocations and their densities are measured using the dislocation analysis method~\cite{stukowski2012automated}. The densities of different types of dislocations during the deformation process are shown in Fig.~\ref{SI9}. In fcc Type 1 (Fig.~\ref{SI9}a) and Type 2 (Fig.~\ref{SI9}b) systems, the predominant dislocations are highly mobile $\frac{1}{6}\langle112\rangle$ (Shockley) partial dislocations and $\frac{1}{6}\langle110\rangle$ (stair-rod) dislocations. In bcc crystals, edge and partial dislocations have high activation energies and thus rarely exist~\cite{hull2011introduction,greer2008comparing}. Screw dislocation is favourable for bcc crystals~\cite{hull2011introduction}. However, these screw dislocations are non-planar and thus exhibit high gliding resistance and low mobility~\cite{hull2011introduction}. This is confirmed by the absence of edge dislocations in Type 3 bcc systems (Fig.~\ref{SI9}c) during the flow stage after yielding ($\varepsilon>6\%$). Therefore, the plastic deformations of bcc composites are mainly in GBs instead of in crystalline grains via dislocation motions.

\begin{figure}[t]
\centering
\includegraphics[width=10cm]{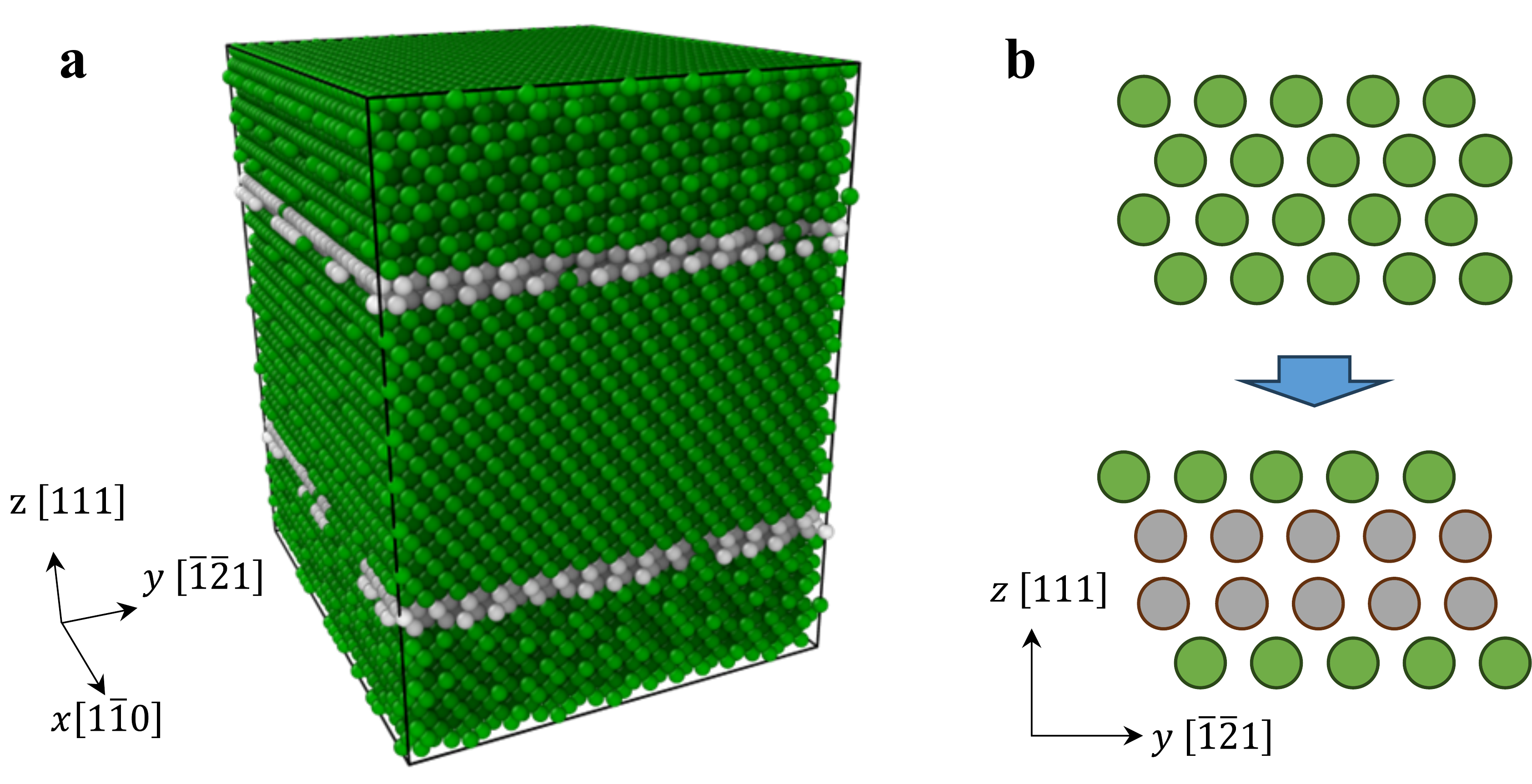}
\caption{\label{SI10} (a) Two stacking faults (grey) along the $(111)$  slip plane in an fcc crystal under periodic boundary conditions. (b) Cross-section on the $y$-$z$ plane showing that a stacking fault can be created by slightly shifting half of a perfect crystal.}
\end{figure}

\subsection{Required stress to create a dislocation in different crystals}

\begin{figure}[h]
\centering
\includegraphics[width=8cm]{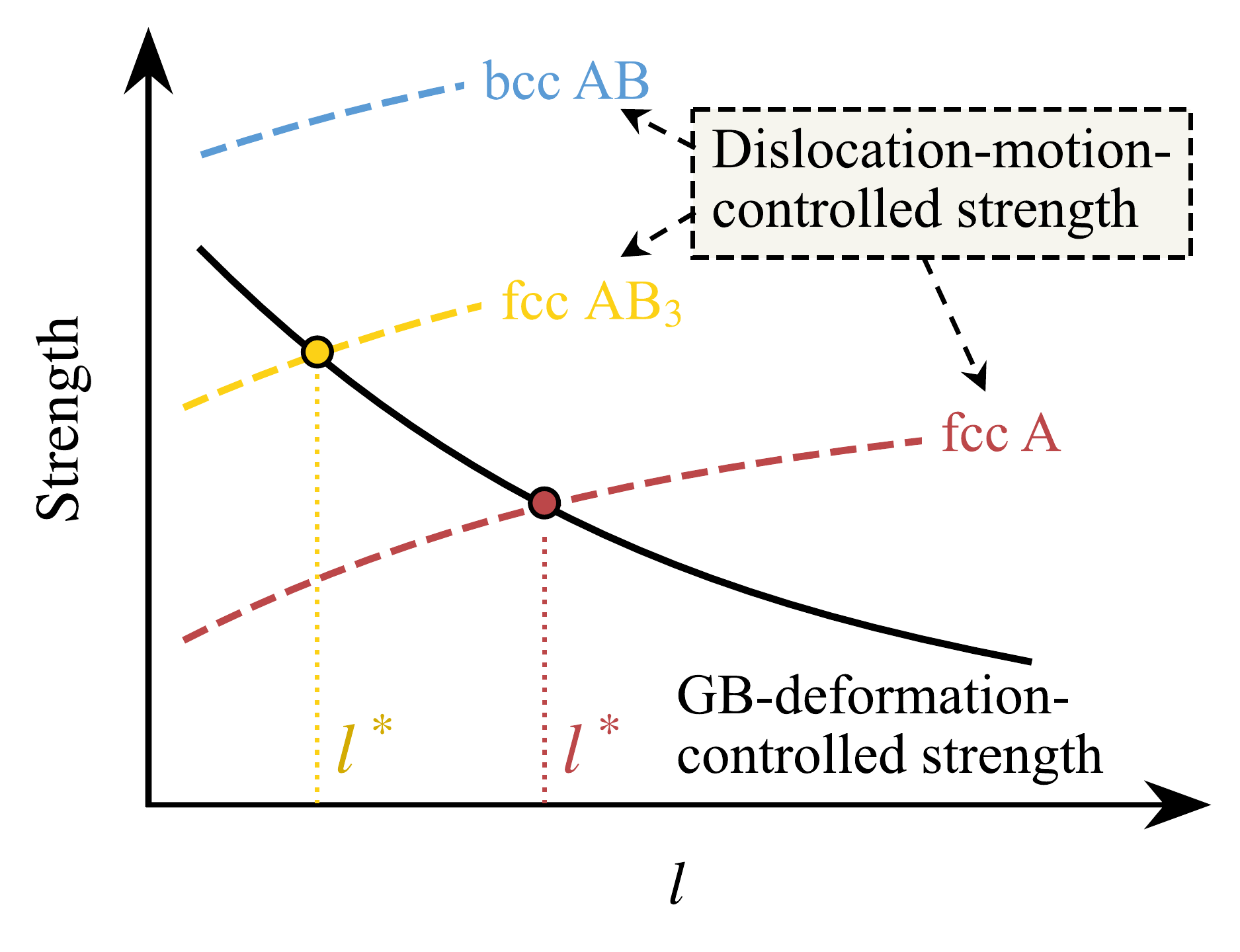}
\caption{\label{SI11} Illustration of solid strength as a function of $l$ under different deformation mechanisms. The dashed curves represent dislocation-motion-controlled strength. The activation stress or friction stress $\sigma_0$ of dislocation motion is much higher in the bcc lattice than in the fcc lattice (Fig.~2k of the main text), thus the dislocation-motion-controlled strength (dashed curves) is high for the bcc lattice. GB deformation is not influenced by different types of lattices, so the GB-controlled strength curve does not shift.}
\end{figure}

The energy of a stacking fault in Fig.~2i of the main text is measured as the potential energy difference between two single crystals with and without the stacking fault~\cite{van2004stacking}. The stacking fault is shown in Fig.~\ref{SI10} and produced as follows.  After an energy minimisation process under $10P^*$, we displace the upper part of the defect-free crystal slightly along the $[\Bar{1}\Bar{2}1]$ direction and set particles to be movable only in the $z$ direction (Fig.~\ref{SI10}), followed by another energy minimisation. 

We compare the critical stresses of creating a dislocation in three types of defect-free single crystals under the same temperature and pressure. Given that a defect-free crystal yields right after the formation of the first dislocation, the yield stress of a defect-free crystal (i.e. maximum of each $\sigma(\varepsilon)$ in Fig.~2j of the main text) is used as the stress required for the generation of dislocation~\cite{salehinia2014crystal}. We observe that a screw dislocation is always generated first in the bcc crystal, and an edge dislocation is always generated first in the fcc crystal at a low stress (Fig.~2j of the main text). These results are consistent with the fact that plastic deformations are dominated by screw dislocations in bcc crystals and by edge dislocations in fcc crystals~\cite{hull2011introduction} because creating a screw dislocation requires much higher stress than that required for edge dislocation in fcc crystals~\cite{hull2011introduction}. 

The activation stress for plastic deformation (i.e. dislocation motions) in crystalline grains depends on the composition and lattice structure, but not on $l$. By contrast, the activation stress for plastic deformation in GBs decreases with $l$ but is independent of the composition and lattice structure of crystalline grains. These results are illustrated in Fig.~2k of the main text. In addition, thick-GB samples contain minimal dislocations, so the dislocation-motion-controlled strength increases with $l$. The combination of this result with those in Fig.~2k provides a sketch of dislocation-motion-controlled strength and GB-deformation-controlled strength as a function of $l$ (shown in Fig.~\ref{SI11}). When one type of strength (e.g. dislocation-motion-controlled) is lower than the other, i.e. this type of deformation is easier, then it will proliferate and largely pre-empt the other type (e.g. GB-controlled) of deformation.

\subsection{Plastic deformations are more uniform in thicker GBs}
\begin{figure}[ht]
\centering
\includegraphics[width=\textwidth]{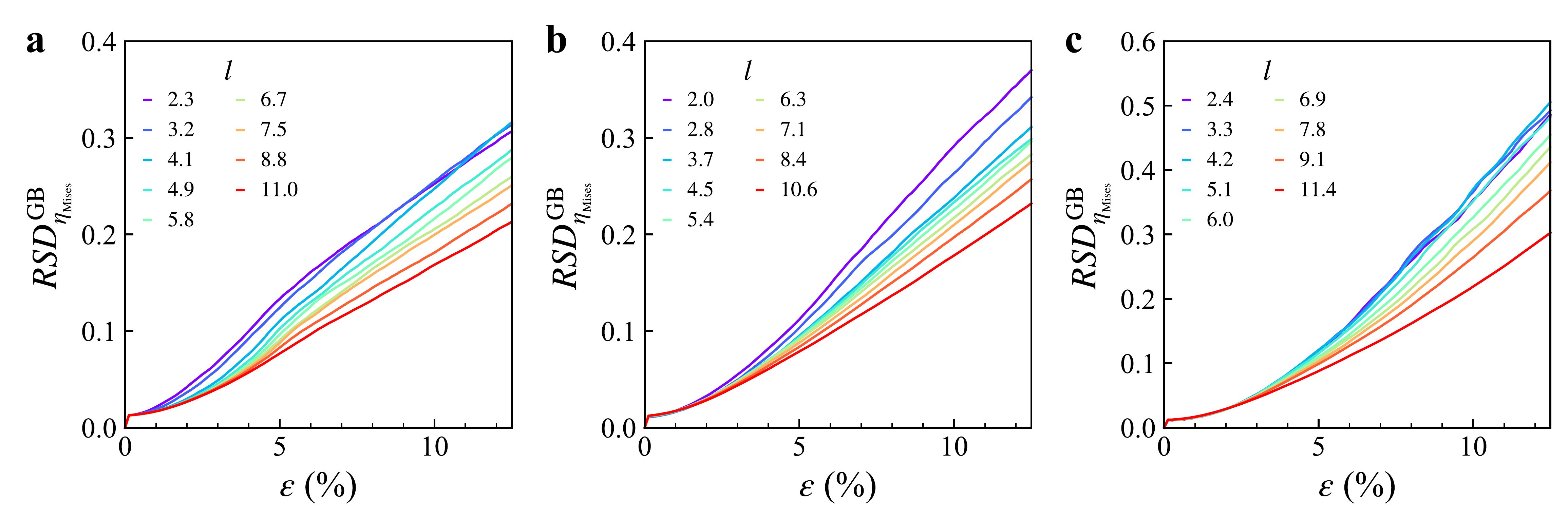}
\caption{\label{SI12} Relative standard deviation (RSD) of von Mises shear strain $\eta_{_{\rm Mises}}$ of particles in GBs during deformation. (a) Type 1 systems with $D=42.05$. (b) Type 2 systems with $D=46.49$. (c) Type 3 systems with $D=46.57$.}
\end{figure}

The uniformity of strain in GBs at the particle length scale can be characterised by the relative standard deviation of $\eta_{_{\rm Mises}}$~\cite{cheng2009correlation}, namely, ${\rm RSD}^{\rm GB}_{\eta_{\rm Mises}}=\sqrt{\sum_i^{N_{\rm GB}}(\eta_{i}-\bar{\eta})^2/N_{\rm GB}}$, where $\eta_{i}$ is the shear strain of particle $i$ and $\bar{\eta}$ is the average over particles in GBs. $N_{\rm GB}$ is the number of particles in GBs. The measured ${\rm RSD}^{\rm GB}_{\eta_{\rm Mises}}$ decreases with $l$ regardless of the composition in the crystalline region (Fig.~\ref{SI12}), indicating that deformations are more uniform at the single-particle scale in thicker GBs.

\section{Fracture process}
Fracture behaviour, whether brittle or ductile, can be identified by fracture morphology. Fractures exhibit various morphologies depending on material properties~\cite{passchier2005microtectonics}, such as ductile fracture featured with necking or shear deformation and brittle facture featured with cracking or brittle shear~\cite{gdoutos2020fracture}. To our knowledge, the fracture process and fracture morphology have not been explored using crystalline--amorphous composites. 

\begin{figure}[htp!]
\centering
\includegraphics[width=15cm]{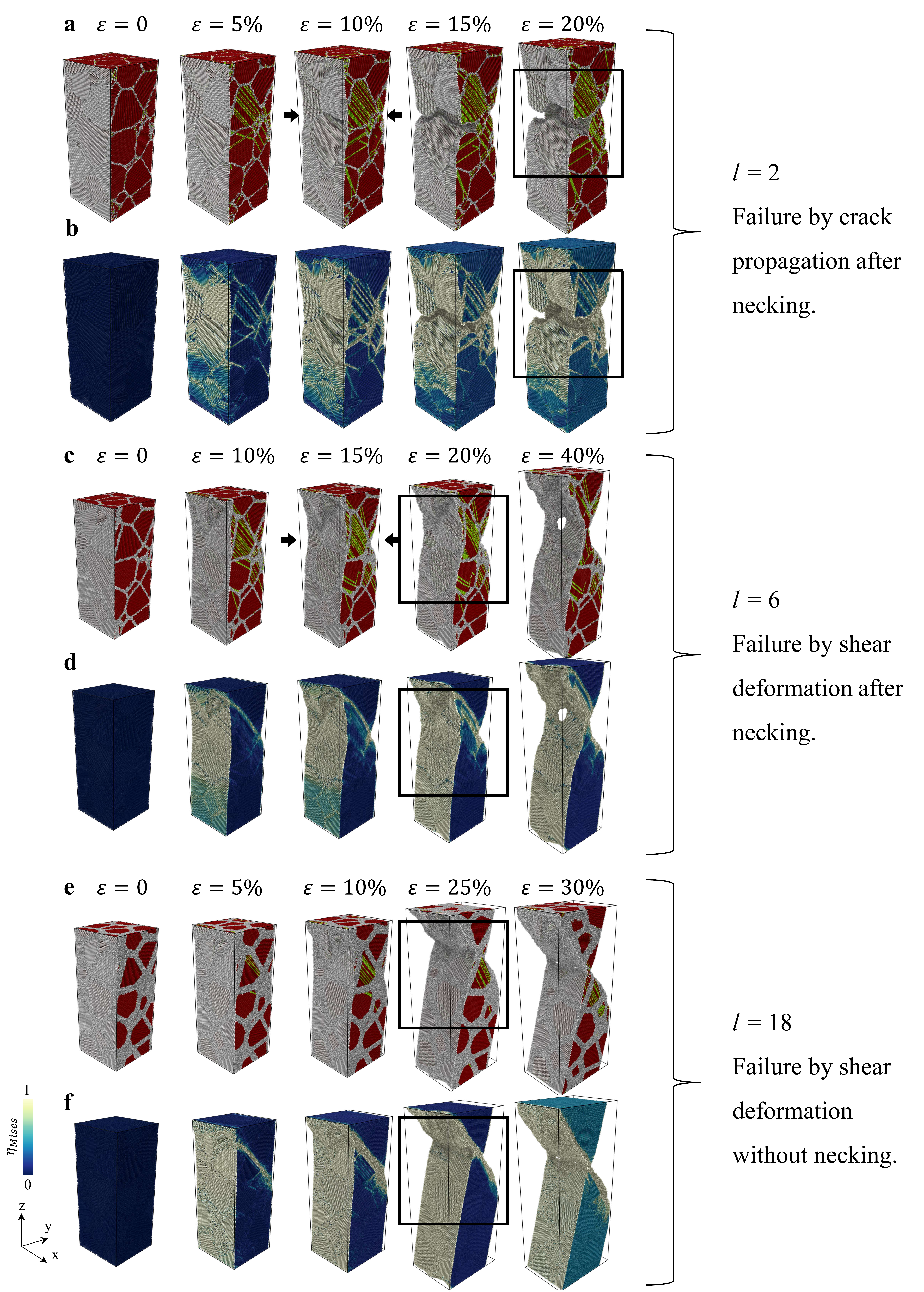}
\caption{\label{SI13} Fracture processes of three fcc samples with $D=45.1$ on the dashed line in Fig.~1c of the main text. (a, b) $l=2.8$, (c, d) $l=6.0$, (e, f) $l=18.0$. Each particle is coloured by its local structure (red: fcc; green: hcp; grey: amorphous) in (a, c, e) and by $\eta_{_{\rm Mises}}$ in (b, d, f). The pairs of black arrows in (a, c) denote necking. The black boxes in (a--f) mark the regions shown in Fig.~3c-h of the main text, respectively.}
\end{figure}

Figure~\ref{SI13} shows different types of ductile fractures in three fcc composites with the same $D$ and different $l$ under the tensile deformation along the $z$ direction. Their stress--strain curves are shown in Fig.~4a of the main text. For thin-GB samples (e.g. $l=2.8$ in Fig.~\ref{SI13}a,b), necking occurs, that is, a prominent decrease in the cross-sectional area occurs. Necking is caused by a local stress concentration with a large amount of strain hardening~\cite{knott1973fundamentals}. We observe that dislocations move into GBs, thus inducing stresses at GBs (Fig.~\ref{SI13}a,b). When $\varepsilon$ reaches 10\%, a crack initiates from a free surface on the $yz$ plane and rapidly propagates along the nearby GBs, resulting in sudden failure at $\varepsilon\simeq20\%$. For the sample with $l=6$ in Fig.~\ref{SI13}c,d, similar dislocation motions induce necking, followed by shear deformation that starts from the necking region, penetrates through GBs and splits the sample (Supplementary Video 5). By contrast, plastic deformation in the thick-GB samples (e.g. $l=18$ in Fig.~\ref{SI13}e,f) occurs solely within thick GBs without obvious necking because few dislocations or stacking faults are produced in crystalline grains during tensile deformation. Thus, GB deformation occurs easily, which pre-empts necking. The shear plane also initiates from a free surface and penetrates through thick GBs, resulting in fracture. 

\begin{figure}[htp!]
\centering
\includegraphics[width=14cm]{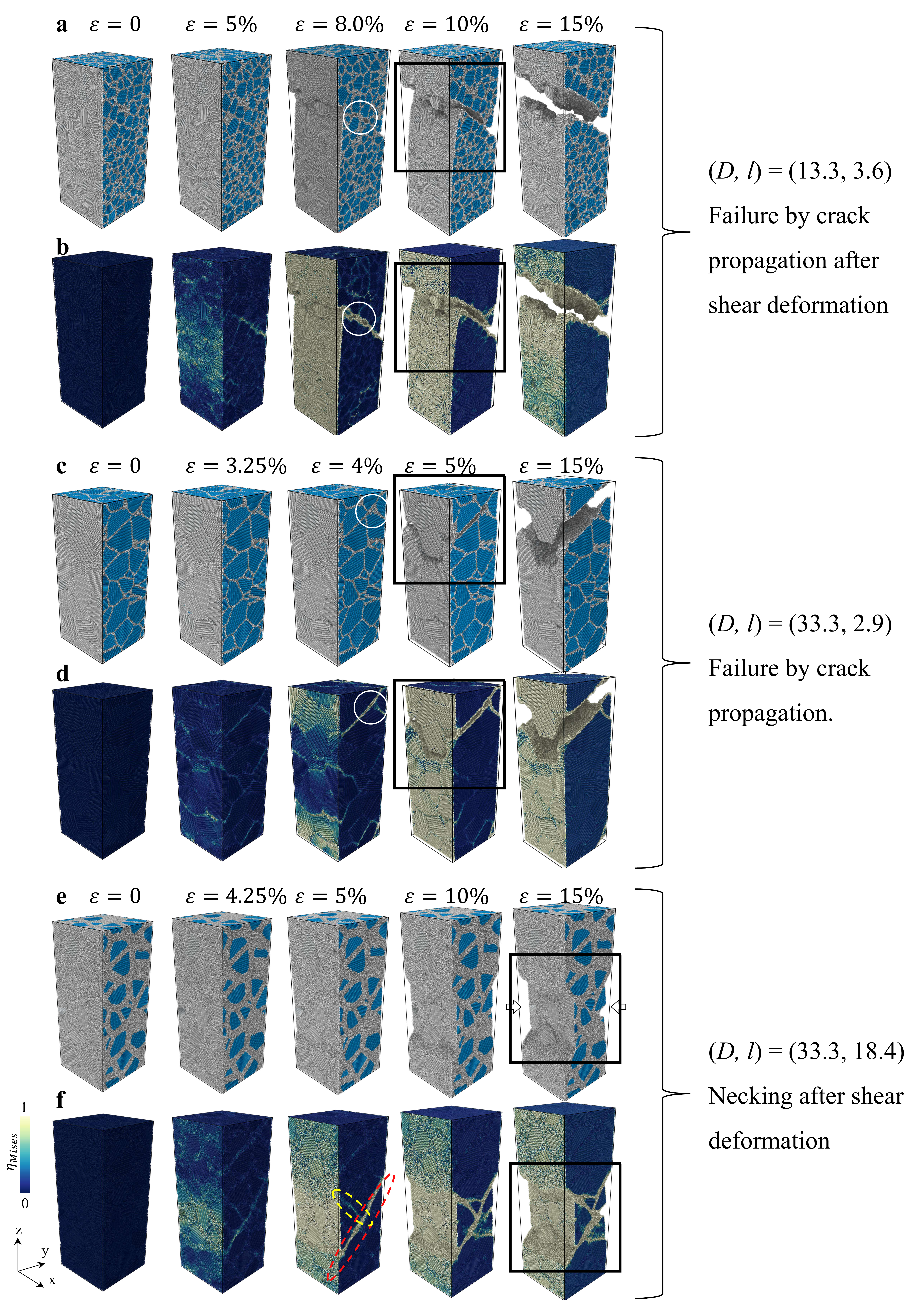}
\caption{\label{SI14} Fracture processes of three bcc samples. (a, b) $(D,l)=(13.3,3.6)$, (c, d) $(D,l)=(33.3,2.9)$, (e, f) $(D,l)=(33.3,18.4)$. Each particle is coloured by its local structure (blue: bcc; grey: amorphous) in (a, c, e) and by $\eta_{_{\rm Mises}}$ in (b, d, f). The red and yellow ellipses in (f) mark two perpendicular shear planes with $\pm45^\circ$ relative to the stretch in the $z$ direction. The white circles in (a--d) mark the voids, which lead to a rough fracture surface. The black boxes in (a--f) mark the regions shown in Fig.~3m-r of the main text, respectively. }
\end{figure}

Figure~\ref{SI14} shows the ductile and brittle fractures of three bcc composites with different $(D,l)$. Their stress--strain curves are shown in Fig.~4k,l of the main text. The deformation process of the bcc polycrystal ($l\simeq 3$) changes from brittle to ductile behaviour as $D$ decreases. The small-$D$ samples undergo ductile shear deformation before $\varepsilon$ reaches $10\%$, and the crack propagates along the GBs and splits the sample, leading to fracture (Fig.~\ref{SI14}a,b). By contrast, the large-$D$ samples in Fig.~\ref{SI14}c,d exhibit no substantial plastic deformation within the grains. When $\varepsilon$ increases to about $4\%$, a crack initiates from a triple junction inside the bulk instead of from a free surface. It rapidly propagates along the GBs, producing a void followed by a rough fracture surface (Fig.~\ref{SI14}c,d). The fracture of the thick-GB bcc samples (Fig.~\ref{SI14}e,f) is similar to that observed in the thick-GB fcc samples (Fig.~\ref{SI13}e,f): plastic deformations occur solely within the thick GBs. Bcc crystals have higher friction stresses for dislocation motions than fcc crystals~\cite{hull2011introduction}, so they have higher resistance to amorphisation~\cite{madec2017dislocation}. This effect hinders the extension of the shear plane and tends to induce another shear plastic plane (yellow ellipse in Fig.~\ref{SI14}f) perpendicular to the initial shear plane (red ellipse in Fig.~\ref{SI14}f), forming necking-like deformation (Fig.~\ref{SI14}e,f). The latter is multi-plane shear or slipping off~\cite{noell2018mechanisms}.

\newpage

\newpage

\section{Supplementary videos}
Supplementary Video 1: Compression processes of two Type 1 systems (fcc A)-(A$_{65}$B$_{35}$). Top row: solid with the maximum strength, i.e. the $\medwhitestar$ in Fig.~1c with $(D,l)=(45.3, 5.8)$. Panels from left to right: stress--strain curve, the system with each particle coloured by its local order (amorphous, fcc or hcp lattice), the system with each particle coloured by its von Mises shear strain and dislocations coloured as Fig.~2d of the main text. Bottom row: results for the solid with $(D,l)=(45.3,11.0)$.\\

Supplementary Video 2: Compression process of two Type 2 systems (fcc AB$_{\rm 3}$)-(A$_{65}$B$_{35}$). Top row: solid with the maximum strength, i.e. the $\medwhitestar$ in Fig.~1f with $(D,l)=(46.9, 2.8)$.  Panels from left to right: stress--strain curve, the system with each particle coloured by its local order (amorphous, fcc or hcp lattice), the system with each particle coloured by its von Mises shear strain and dislocations coloured as Fig.~2e of the main text. Bottom row: results for the solid with $(D,l)=(46.9,10.6)$.\\

Supplementary Video 3: Compression process of two Type 3 systems (bcc AB)-(A$_{65}$B$_{35}$). Top row: solid with the maximum strength, i.e. the $\medwhitestar$ in Fig.~1i with $(D,l)=(46.6, 2.4)$. Panels from left to right: stress--strain curve, the system with each particle coloured by its local order (amorphous, bcc lattice), the system with each particle coloured by its von Mises shear strain and dislocations coloured as Fig.~2f of the main text. Bottom row: results for the solid with $(D,l)=(46.6,11.4)$.\\

Supplementary Video 4: Tensile deformation process of the Type 1 system (fcc A)-(A$_{65}$B$_{35}$) with $(D,l)=(45.1,2.8)$. A snapshot is shown in Fig.~4c,f of the main text. Panels from left to right: stress--strain curve, the system with each particle coloured by its local order (amorphous, fcc, or hcp lattice) and the system with each particle coloured by its von Mises shear strain.\\

Supplementary Video 5: Tensile deformation process of the Type 1 system (fcc A)-(A$_{65}$B$_{35}$) with $(D,l)=(45.1,6.0)$. A snapshot is shown in Fig.~4d,g of the main text. Panels from left to right: stress--strain curve, the system with each particle coloured by its local order (amorphous, fcc, or hcp lattice) and the system with each particle coloured by its von Mises shear strain.\\

Supplementary Video 6: Tensile deformation process of the Type 1 system (fcc A)-(A$_{65}$B$_{35}$) with $(D,l)=(45.1,18)$. A snapshot is shown in Fig.~4e,h of the main text. Panels from left to right: stress--strain curve, the system with each particle coloured by its local order (amorphous, fcc, or hcp lattice) and the system with each particle coloured by its von Mises shear strain.\\

Supplementary Video 7: Tensile deformation process of the Type 3 system (bcc AB)-(A$_{65}$B$_{35}$) with $(D,l)=(13.2,3.6)$. A snapshot is shown in Fig.~4m,p of the main text. Panels from left to right: stress--strain curve, the system with each particle coloured by its local order (amorphous or bcc lattice) and the system with each particle coloured by its von Mises shear strain.\\

Supplementary Video 8: Tensile deformation process of the Type 3 system (bcc AB)-(A$_{65}$B$_{35}$) with $(D,l)=(33.3,2.9)$. A snapshot is shown in Fig.~4n,q of the main text. Panels from left to right: stress--strain curve, the system with each particle coloured by its local order (amorphous or bcc lattice) and the system with each particle coloured by its von Mises shear strain.\\

Supplementary Video 9: Tensile deformation process of the Type 3 system (bcc AB)-(A$_{65}$B$_{35}$) with $(D,l)=(33.3,18.4)$. A snapshot is shown in Fig.~4o,r of the main text. Panels from left to right: stress--strain curve, the system with each particle coloured by its local order (amorphous or bcc lattice), and the system with each particle coloured by its von Mises shear strain.
\newpage

%\bibliographystyle{naturemag}
%\bibliography{SI-bibliography}
%\bibliography{extractred_bib_SI}